\documentclass{emulateapj}

\usepackage{verbatim}
\usepackage{amsmath, amsthm, amssymb}
\usepackage{times}
\usepackage{graphics}
\usepackage{graphicx}
\usepackage{rotate}
\usepackage{epsfig}
\usepackage{latexsym} 
\usepackage{wrapfig}
\usepackage{natbib}
\usepackage{setspace}
\usepackage{epstopdf}
\usepackage{chngcntr}
\usepackage{paralist}
\usepackage[T1]{fontenc} 

\def\alf{Alfv\'en}
\def\ga{\mathrel{\hbox{\rlap{\hbox{\lower4pt\hbox{$\sim$}}}\hbox{$>$}}}}
\def\la{\mathrel{\hbox{\rlap{\hbox{\lower4pt\hbox{$\sim$}}}\hbox{$<$}}}}

\shorttitle{Synchrotron Fluctuation Statistics}

\shortauthors{HERRON ET AL. }

\begin{document}

\title{Radio synchrotron fluctuation statistics as a probe of magnetized interstellar turbulence}

\author{C. A. Herron\altaffilmark{1},  Blakesley Burkhart\altaffilmark{2}, A. Lazarian\altaffilmark{3}, B. M. Gaensler\altaffilmark{4, 1}, N. M. McClure-Griffiths\altaffilmark{5} }
\altaffiltext{1}{Sydney Institute for Astronomy, School of Physics, University of Sydney, New South Wales 2006, Australia; C.Herron@physics.usyd.edu.au}
\altaffiltext{2}{Harvard-Smithsonian Center for Astrophysics, 60 Garden St. Cambridge, MA, USA}
\altaffiltext{3}{Astronomy Department, University of Wisconsin, Madison, 475 N. 
Charter St., WI 53711, USA}
\altaffiltext{4}{Dunlap Institute for Astronomy and Astrophysics, University of Toronto, 50 St. George Street, Toronto, Ontario, M5S 3H4, Canada}
\altaffiltext{5}{Research School of Astronomy and Astrophysics, The Australian National University, Canberra, ACT 2611, Australia}

\begin{abstract}

We investigate how observations of synchrotron intensity fluctuations can be used to probe the sonic and \alf ic Mach numbers of interstellar turbulence, based on mock observations performed on simulations of magnetohydrodynamic turbulence. We find that the structure function slope, and a diagnostic of anisotropy that we call the integrated quadrupole ratio modulus, both depend on the \alf ic Mach number. However, these statistics also depend on the orientation of the mean magnetic field in the synchrotron emitting region relative to our line of sight, and this creates a degeneracy that cannot be broken by observations of synchrotron intensity alone. We conclude that the polarization of synchrotron emission could be analyzed to break this degeneracy, and suggest that this will be possible with the Square Kilometre Array.

\end{abstract}

\section{Introduction}
\label{intro}

Turbulence and magnetic fields are both ubiquitous in the Milky Way \citep{Ferriere2001, Elmegreen2004, Falceta-Goncalves2014, Haverkorn2015}, and significantly affect the formation of stars \citep{McKee2007, Dobbs2014}, how gas is recycled in the Galaxy, dynamo amplification of the Galactic magnetic field \citep{Beresnyak2015, Brandenburg2015}, and cosmic ray propagation \citep{Scalo2004}. However, the nature of turbulent flows in the magnetized interstellar medium (ISM) is largely unknown, as key properties of turbulence are poorly constrained by observations. Thus, to better understand these galactic processes, and galaxy evolution as a whole, we require methods to quantitatively analyze turbulence in the ISM. 

Due to the random nature of turbulence, these methods must examine statistics of observable quantities that depend upon either the density, velocity or magnetic field of the interstellar medium to recover the properties of turbulence in the observed ISM phase. These properties include the sonic and \alf ic Mach numbers,
\begin{equation}
\mathcal{M}_s = \left< \frac{|\mathbf{v}|}{c_s} \right>, \quad \text{and} \quad \mathcal{M}_A = \left< \frac{|\mathbf{v}|}{v_A} \right> , \label{Mach_nums}
\end{equation}
respectively, where $\mathbf{v}$ is the velocity vector of a parcel of gas, $c_s$ is the speed of sound, and $v_A = \frac{|\mathbf{B}|}{\sqrt{\rho}}$ is the \alf \, speed, with $\mathbf{B}$ the magnetic field in the parcel of gas, and $\rho$ its density. These Mach numbers describe how fast the gas motions are relative to the sound speed and \alf \, speed respectively. 

For example, turbulence in the warm neutral medium of the ISM can be studied by observing the 21 cm emission from neutral hydrogen. Analysis of these velocity statistics of neutral hydrogen in the Milky Way with the Velocity Channel Analysis and Velocity Correlation Spectrum techniques (\citealt{Lazarian2000, Lazarian2006}, see \citealt{Lazarian2009} for a review) has enabled the velocity spectrum of the warm neutral medium to be studied (for example, \citealt{Chepurnov2010}). Additionally, \cite{Burkhart2010} employed the third and fourth order statistical moments of the neutral hydrogen column density of the Small Magellanic Cloud to show that approximately $90 \%$ of its neutral gas is sub- or transonic.

Another physical process that can be used to study interstellar turbulence is linearly polarized synchrotron radiation, emitted by ultra-relativistic electrons spiralling around magnetic field lines. Recently, \cite{Gaensler2011} and \cite{Burkhart2012} have shown that the skewness and kurtosis of polarization gradients of synchrotron emission can constrain the sonic Mach number in the warm ionized medium (WIM). This technique has been applied by these authors and \cite{Iacobelli2014} to observations, all of which demonstrate that the WIM is approximately transonic. 

In this paper, we investigate how observed statistics of the total intensity of synchrotron radiation can be used to study turbulence in the ISM. To determine the intensity of the synchrotron radiation, we assume that the ultra-relativistic electrons have a homogeneous and isotropic power-law energy distribution of the form

\begin{equation}
N(E) \, \mathrm{d} E = K E^{2 \alpha - 1} \, \mathrm{d} E, \label{energy_dist}
\end{equation}
where $N(E)$ is the number density of electrons with energy between $E$ and $E+\mathrm{d}E$, $K$ is a normalization constant and $\alpha$ is the spectral index. The emitted synchrotron intensity $I$, at frequency $\nu$, is then given by \citep{Ginzburg1965}:

\begin{align}
I(\nu) &= \frac{e^3}{4 \pi m_e c^2} \int_0^{L} \frac{\sqrt{3}}{2 - 2 \alpha} \Gamma \left( \frac{2 - 6 \alpha}{12} \right) \Gamma \left( \frac{22 - 6 \alpha}{12} \right) \times \nonumber \\ 
& \quad \quad \left( \frac{3e}{2 \pi m_e^3 c^5} \right)^{-\alpha} K B_{\perp}^{\gamma} \nu^{\alpha} \, \mathrm{d} L \label{sync_inten}
\end{align}
where $e$ and $m_e$ are the charge and mass of an electron, $c$ is the speed of light, $\gamma = 1 - \alpha$ is the exponent of $B_{\perp}$, the strength of the magnetic field perpendicular to the line of sight, $\Gamma$ denotes the gamma function, and $\mathrm{d} L$ represents an infinitesimal length along the line of sight. The integral runs from the back of the volume of synchrotron-emitting electrons to the observer at $L=0$.

Early studies on how to use statistics of synchrotron intensity to study turbulent magnetic fields either assumed that $\gamma = 2$, or that the magneto-hydrodynamic (MHD) turbulence was isotropic \citep{Chibisov1981, Lazarian1990, Lazarian1991, Chepurnov1998}. However, there is theoretical (\citealt{Goldreich1995}, see review by \citealt{Brandenburg2013}) and numerical (\citealt{Shebalin1983, Cho2002}, see review by \citealt{Cho2003a}) evidence that MHD turbulence should be anisotropic, with the plasma flows elongated in the direction of the local magnetic field. This causes the synchrotron intensity images to appear to have filaments, corresponding to the stronger magnetic field strength in the plasma flows. There are also theoretical reasons why $\gamma$ should not be constant, but vary throughout a galaxy. Immediately after electrons are first accelerated to ultra-relativistic speeds, which may occur by diffusive shock acceleration in a supernova explosion, the cosmic ray electron population typically creates a radio spectrum with index between $-0.1$ and $-0.5$, corresponding to $1.1 < \gamma < 1.5$ \citep{Green1988}. As the ultra-relativistic electrons propagate, they lose energy via their emitted synchrotron radiation, inverse-Compton scattering, and bremsstrahlung. These energy losses affect high energy electrons more than low energy electrons, causing the synchrotron spectrum to steepen, resulting in more negative values of $\alpha$, and larger values of $\gamma$. An older, steep spectrum electron population has diffused away from the injection point, and so occupies a greater volume than the young, flat spectrum electron population. Hence, it is expected that we should find flat-spectrum synchrotron emission near star-forming regions in the Galactic plane, but steep-spectrum emission further from the plane. This is supported by observations; for example, \cite{Bennett2003} found that $\gamma$ varies with Galactic latitude, being smallest in star-forming regions ($1.5$) and larger in the halo ($2.1$), and \cite{Tabatabaei2008} and \cite{Basu2015} found that in nearby spiral galaxies, the diffuse synchrotron emission has a flatter spectrum in the spiral arms than in the inter-arm regions. Although values of $\gamma$ between $1.5$ and $2.5$ are more typical of synchrotron radiation in the Milky Way, in this paper we will analyze values of $\gamma$ between $1$ and $4$, to cover all values likely to be observed.

Recently, \cite{Lazarian2012}, hereafter LP12, studied how the statistics of synchrotron emission are related to the underlying turbulent magnetic field analytically, without assuming that $\gamma = 2$, or isotropic MHD turbulence. They predicted that the normalized correlation function (NCF) of synchrotron intensity is insensitive to $\gamma$, and hence is an ideal statistic to be applied to observations, because it provides a more robust link between the statistics of synchrotron intensity and the statistics of the magnetic field. LP12 also introduced a new statistic for quantitatively describing the observed degree of anisotropy in MHD turbulence, that we will call the quadrupole ratio, and suggested that this statistic could constrain the compressibility of a plasma, as well as the orientation of the mean magnetic field relative to the line of sight. 

In this paper, we calculate the NCFs of mock synchrotron intensity maps generated from simulations of ideal MHD turbulence, to test the prediction that this statistic is insensitive to $\gamma$. Additionally, we investigate how the NCF and the quadrupole ratio are related to the properties of the underlying magnetized turbulence, in the hope of constructing a technique that allows us to characterize magnetized turbulence in the ISM from observations of diffuse synchrotron emission. Such a technique may involve plotting each simulated synchrotron statistic against the sonic or \alf ic Mach number, for instance, and using the corresponding range of Mach numbers compatible with an observed statistic to constrain each parameter of the turbulence in the data. Our ability to estimate parameters of turbulence is then determined by the observational uncertainty in the measured statistic, the uncertainty in the simulated statistics due to random fluctuations in the turbulence, and our ability to resolve any degeneracy between parameters of turbulence.

This study complements recent analyses of total synchrotron intensity by \cite{Iacobelli2013} and \cite{Stepanov2014}. \cite{Iacobelli2013} used fluctuations in the total synchrotron intensity to constrain the outer scale of turbulence in the Fan region, and to show that the ratio of the random to ordered magnetic field strengths changes with Galactic coordinates, suggesting different turbulent regimes. \cite{Stepanov2014} showed that cosmic rays are not in local energy equipartition with magnetic fields, and from this, that the distribution of cosmic rays in the Milky Way is uniform on scales of at least 100 pc.

In Section \ref{theory} we introduce the analytic theory of synchrotron fluctuations developed by LP12, and the results pertinent to this paper. In Section \ref{method} we introduce the simulations of ideal MHD turbulence that we use, the method that we use to construct mock synchrotron intensity maps, and the properties of turbulence we examine. In Section \ref{results} we test the predictions of LP12, and investigate how the statistics of synchrotron fluctuations depend on the properties of the turbulence and the orientation of the mean magnetic field relative to the line of sight. We discuss the possibility of determining properties of turbulence from observations of diffuse synchrotron emission in Section \ref{discuss}, and conclude in Section \ref{conclusion}.

\section{Theory of Synchrotron Fluctuations}
\label{theory}

For a fixed observing frequency, Eq. \ref{sync_inten} demonstrates that fluctuations in synchrotron intensity are caused by fluctuations in the strength of the magnetic field perpendicular to the line of sight, $\gamma$, $K$, and path length variations. In general, the first three of these variables will vary over the sky and along the line of sight. $B_{\perp}$ will vary due to turbulent fluctuations in the Galactic magnetic field and the relative orientation between the line of sight and the magnetic field. $\gamma$ and $K$ will vary due to the cosmic rays having different ages or input particle spectra. In this paper, we will be considering the synchrotron intensity observed over small fields of view, where we assume that all lines of sight through the emitting region have the same length, and that $\gamma$ and $K$ are constant throughout the emitting region. These latter assumptions are valid because $\gamma$ varies on larger scales in the Milky Way than the scales we consider, and \cite{Stepanov2014} showed that the distribution of cosmic rays, and hence $K$, is uniform on scales of at least $100$ pc. The consequence of these assumptions is that within any small field of view, the fluctuations in synchrotron intensity are only caused by fluctuations in the magnetic field. 

However, as $\gamma$ and $K$ can differ between fields of view, we require statistics to be independent of $\gamma$ and $K$, so that variations in the statistics of synchrotron intensity calculated across the sky are only caused by changes in the alignment and amplitude distribution of the projected magnetic field. Such statistics of synchrotron intensity are sensitive only to the parameters of magnetized turbulence, such as the sonic Mach number, and hence provide the simplest means by which to quantitatively determine the values of these parameters. As $K$ is assumed to be constant within any given field of view, it only serves to scale the observed synchrotron intensity by some factor, and hence should not influence any of the statistics calculated in this paper. Thus, we will only examine the $\gamma$ dependence of statistics henceforth. 

One potentially useful statistic is the NCF of synchrotron emissivity:

\begin{equation}
\xi_{B_{\perp}^{\gamma}}(\mathbf{r}) = \frac{\langle B_{\perp}^{\gamma}(\mathbf{x}) B_{\perp}^{\gamma}(\mathbf{x}+\mathbf{r}) \rangle - \langle B_{\perp}^{\gamma}(\mathbf{x}) \rangle^2}{\langle B_{\perp}^{\gamma}(\mathbf{x})^2 \rangle - \langle B_{\perp}^{\gamma}(\mathbf{x}) \rangle^2}, \label{norm_corr_emiss}
\end{equation}
where $\mathbf{x} = (x,y,z)$ is a position vector to a point in the emitting region, described by a Cartesian coordinate system in which the z-axis is along the line of sight. $\mathbf{r}$ is the separation vector between two points being used to calculate the correlation function, and $\langle ... \rangle$ denotes an average over all $\mathbf{x}$. Under the assumption of an isotropic Gaussian field, LP12 found that $\xi_{B_{\perp}^{\gamma}}$ only has a weak dependence on $\gamma$. Furthermore, by applying these equations to an analytic model of turbulence involving a mean magnetic field along the x axis, with fluctuations only along the y axis, LP12 found that the NCF of synchrotron emissivity weakly depends on $\gamma$ in this anisotropic case, and becomes more sensitive to $\gamma$ as the mean magnetic field strength decreases.

We can also define the normalized correlation coefficients for the components of the magnetic field perpendicular to the line of sight, i.e. $B_x$ and $B_y$, by

\begin{align}
c_1(\mathbf{r}) &= \frac{\langle B_x(\mathbf{x}) B_x(\mathbf{x}+\mathbf{r}) \rangle - \langle B_x(\mathbf{x}) \rangle^2}{\langle B_x(\mathbf{x})^2 \rangle - \langle B_x(\mathbf{x}) \rangle^2}  \label{c_1} \\
c_2(\mathbf{r}) &= \frac{\langle B_y(\mathbf{x}) B_y(\mathbf{x}+\mathbf{r}) \rangle - \langle B_y(\mathbf{x}) \rangle^2}{\langle B_y(\mathbf{x})^2 \rangle - \langle B_y(\mathbf{x}) \rangle^2},  \label{c_2}
\end{align}
respectively. Based on their finding that the NCF of synchrotron emissivity is insensitive to $\gamma$, LP12 predicted that for an isotropic Gaussian magnetic field, and any $\gamma$ values between $1$ and $4$, $\xi_{B_{\perp}^{\gamma}}$ can be approximated by the NCF for $\gamma = 2$, namely

\begin{equation}
\xi_{B_{\perp}^{\gamma}} \approx \frac{1}{2} (c_1^2 + c_2^2).  \label{sync_emis_approx}
\end{equation}

The corresponding equation for anisotropic turbulence is given by Eq. 28 in LP12, but for simplicity we will only test the equation for the case of isotropic turbulence in this paper. Following these results, LP12 conjecture that the NCF of the observed synchrotron intensity, given by

\begin{equation}
\xi_{I}(\mathbf{R}) = \frac{\langle I(\mathbf{X}) I(\mathbf{X}+\mathbf{R}) \rangle - \langle I(\mathbf{X}) \rangle^2}{\langle I(\mathbf{X})^2 \rangle - \langle I(\mathbf{X}) \rangle^2}, \label{norm_corr_inten}
\end{equation}
for a vector $\mathbf{X} = (x,y)$ describing the location of a point on the plane of the sky, and for a separation vector in the plane of the sky $\mathbf{R}$, is also insensitive to $\gamma$, and show that this is true for an isotropic random field. LP12 also explain that the structure function of synchrotron intensity, given by

\begin{equation}
D_I(\mathbf{R}) = \langle (I(\mathbf{X}) - I(\mathbf{X} + \mathbf{R}))^2 \rangle, \label{sync_SF}
\end{equation}
is the preferred statistic to use when analyzing synchrotron fluctuations, because it is linear on small scales, unlike the correlation function. It is possible to calculate a normalized structure function for synchrotron intensity from the NCF, by using the formula

\begin{equation}
\tilde{D}_I = 2(1 - \xi_I). \label{norm_SF}
\end{equation}
As the NCF is expected to only weakly depend on $\gamma$, it is hence expected that the normalized structure function should also weakly depend on $\gamma$.

Conventionally, the power spectrum of an image of the column density of a turbulent volume of gas is said to be proportional to $k^{-(3+m)}$, where $k$ is the wavenumber, and $m$ is a constant that helps to identify the type of turbulence. For example, $m=2/3$ corresponds to Kolmogorov turbulence. Under this convention, the corresponding structure function is proportional to $R^{1+m}$. We will be using $m$ to characterize the slope of the structure function throughout this paper.

Synchrotron intensity images can also be characterized by how anisotropic they are, where the anisotropy could be introduced by the interstellar magnetic field constraining ionized gas into filaments. To measure this anisotropy, LP12 suggested using the multipole moments of the normalized structure function, given by

\begin{equation}
\tilde{M}_n(R) = \frac{1}{2 \pi} \int_0^{2 \pi} e^{-i n \phi} \tilde{D}_I(R, \phi) \, \mathrm{d} \phi . \label{multipole}
\end{equation}

In Eq. \ref{multipole}, $n$ determines the order of the multipole to calculate, $R$ represents the radial distance between the points being used to calculate the structure function, and $\phi$ is the polar angle subtended by the horizontal and a line connecting the points, both measured in the plane of the sky. The $n=0$ term gives the monopole, which measures the isotropic part of the normalized structure function, and the $n=1$ term gives the dipole, which is identically zero because of the rotational symmetry of the structure function. Thus, the lowest order, non-zero multipole describing the anisotropy of the normalized structure function is the $n=2$ term, which gives the quadrupole moment. 

To produce a more robust statistic of anisotropy, LP12 suggested normalizing the quadrupole moment by dividing by the monopole moment. By using the rotational symmetry of the normalized structure function, this statistic can be written as

\begin{align}
\frac{\tilde{M}_2(R)}{\tilde{M}_0(R)} &= \frac{2 \int_0^{\pi} \cos (2 \phi) \tilde{D}_I(R, \phi) \, \mathrm{d} \phi}{\int_0^{2 \pi} \tilde{D}_I(R, \phi) \, \mathrm{d} \phi} \qquad \qquad \qquad \qquad \nonumber \\
& \qquad \qquad \quad - \frac{2 i \int_0^{\pi} \sin(2 \phi) \tilde{D}_I(R, \phi) \, \mathrm{d} \phi}{\int_0^{2 \pi} \tilde{D}_I(R, \phi) \, \mathrm{d} \phi}. \label{quad_ratio}
\end{align} 

The real part of $\tilde{M}_2(R)/\tilde{M}_0(R)$ is sensitive to anisotropy along the vertical and horizontal axes of the image, and the imaginary part is sensitive to anisotropy along the diagonals of the image. Thus, the modulus of $\tilde{M}_2(R)/\tilde{M}_0(R)$ provides a full characterization of the amplitude of any anisotropy (i.e. the filamentarity of the synchrotron intensity), and the argument describes the orientation of the axis of anisotropy, i.e. in what direction filaments of synchrotron intensity are elongated. Both of these quantities provide useful diagnostics of the observed turbulence. However, as the argument of $\tilde{M}_2(R)/\tilde{M}_0(R)$ mainly describes the direction of the mean magnetic field projected onto the plane of the sky, which can alternatively be determined by using the polarization angle of synchrotron emission observed at high frequencies, we will focus on the modulus for the remainder of this paper. We will henceforth refer to the modulus of $\tilde{M}_2(R)/\tilde{M}_0(R)$ as the quadrupole ratio modulus, and to the real and imaginary parts of $\tilde{M}_2(R)/\tilde{M}_0(R)$ as the real and imaginary parts of the quadrupole ratio.

LP12 estimated that the quadrupole ratio modulus should be approximately $0.4$ for sub-\alf ic turbulence (\alf ic Mach number less than $1$, gas dynamics dominated by the magnetic field), and $0.3$ for trans-\alf ic turbulence (\alf ic Mach number approximately 1). For turbulence in the super-\alf ic regime (\alf ic Mach number greater than $1$), the quadrupole ratio modulus was estimated to be approximately $0.1$. These estimates are influenced by the assumed model of magnetized turbulence.

In Section \ref{cfClaim} we will test the prediction that the NCF of synchrotron emissivity is insensitive to $\gamma$, by calculating this quantity for simulations of MHD turbulence. We will also test whether the NCF can be approximated by the NCF for $\gamma = 2$, and the conjecture that the NCF of synchrotron intensity is also insensitive to $\gamma$. In Section \ref{quadClaim}, we test whether the quadrupole ratio modulus is related to the \alf ic Mach number as predicted by LP12.

\section{Methodology}
\label{method}

To test the predictions of the synchrotron fluctuation theory developed by LP12, as well as the suitability of the structure function and quadrupole ratio modulus as tracers of properties of turbulence, we will apply the theory to numerical simulations of MHD turbulence. In this section we will discuss how our simulations of MHD turbulence were performed, and how mock synchrotron intensity maps were produced for the simulations.

We use the second-order-accurate hybrid essentially non-oscillatory code written by \cite{Cho2003} to numerically solve the ideal MHD equations in a periodic box, namely:

\begin{equation}
\frac{\partial \rho}{\partial t} + \nabla \cdot (\rho \mathbf{v}) = 0, \label{mass_cons}
\end{equation}

\begin{equation}
\rho \left[ \frac{\partial \mathbf{v}}{\partial t} + (\mathbf{v} \cdot \nabla) \mathbf{v} \right] +  \nabla p - \frac{1}{4 \pi} \mathbf{J} \times \mathbf{B} = \mathbf{f}, \label{energy_cons}
\end{equation}

\begin{equation}
\frac{\partial \mathbf{B}}{\partial t} - \nabla \times (\mathbf{v} \times \mathbf{B}) = 0, \label{Faraday}
\end{equation}

\begin{equation}
\nabla \cdot \mathbf{B} = 0. \label{Gauss}
\end{equation}

In Equations \ref{mass_cons} to \ref{Gauss}, $p$ is the thermal gas pressure, $\mathbf{J} = \nabla \times \mathbf{B}$ is the current density, $t$ is time, and $\mathbf{f}$ is a driving force. Each of these quantities are normalized, so that Equations \ref{mass_cons} to \ref{Gauss} become dimensionless, and the simulations are scale-free. The quantities are hence measured in simulation units. 

\begin{figure}
\centering
\includegraphics[scale=0.4]{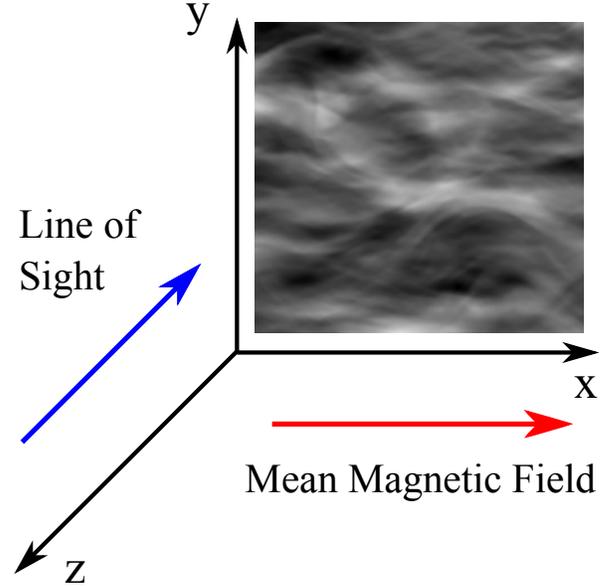}
\caption{The initial set-up of the simulation cube. The mean magnetic field is along the x-axis, and the line of sight is along the z-axis. Maps of synchrotron intensity are produced by integrating the synchrotron emissivity along the z-axis. An example synchrotron intensity map is shown inset, produced with the Ms0.48Ma0.65 simulation. In the grayscale, white represents intense synchrotron emission, and black represents faint synchrotron emission.} \label{fig1}
\end{figure}

In our simulations we drive turbulence by setting $\mathbf{f}$ to be a random solenoidal driving force that acts on large scales. We also use an isothermal equation of state, $p = c_s^2 \rho$, so that we simulate compressible turbulence. Each simulation run starts with fully ionized gas of uniform pressure and unit density in a cube that has $512$ pixels along each side, and a uniform magnetic field along the x-axis. We consider lines of sight that are perpendicular to the mean field, oriented along the y- or z-axes, as shown in Figure \ref{fig1}. 

The free variables in each simulation run are the initial pressure, which controls the sonic Mach number, and the initial magnetic field strength, which controls the \alf ic Mach number. These two variables were adjusted to run $18$ simulations, as listed in Table \ref{TabSims}, which cover different regimes of turbulence. Each simulation is assigned a code of the form Ms0.45Ma1.72, for example, which means that the sonic Mach number is 0.45, and the \alf ic Mach number is 1.72. Of these simulations, we expect that the Ms0.87Ma0.7 and Ms0.48Ma0.65 simulations are most similar to the WIM of the Milky Way, because they are sub- or transonic, and the strength of the random component of the magnetic field in these simulations is comparable to the strength of the uniform field, as found in the Milky Way (\citealt{Gaensler2011}, \citealt{Iacobelli2014}, \citealt{Sun2008}).

\begin{table*}
\centering
\caption{Parameters used to run the different MHD turbulence simulations, and the sonic and Alfvenic Mach numbers of each simulation for the snapshot used.} \label{TabSims}
\begin{tabular}{| c c | c c | c c | c |}
\hline
\hline
Sim No. & Code & Init B (sim units) & Init P (sim units) & $M_s$ & $M_A$ & Turbulence Regime\\
\hline
1 & Ms10.96Ma1.41  & $0.1$ & $0.0049$ & $10.96$ & $1.41$ & Supersonic and trans-\alf ic \\
2 & Ms9.16Ma1.77  & $0.1$ & $0.0077$ & $9.16$ & $1.77$ & Supersonic and super-\alf ic \\
3 & Ms7.02Ma1.76  & $0.1$ & $0.01$     & $7.02$ & $1.76$ & \textquotedbl \\
4 & Ms4.32Ma1.51  & $0.1$ & $0.025$   & $4.32$ & $1.51$ & \textquotedbl \\
5 & Ms3.11Ma1.69  & $0.1$ & $0.05$     & $3.11$ & $1.69$ & \textquotedbl \\
6 & Ms2.38Ma1.86  & $0.1$ & $0.1$    & $2.38$ & $1.86$ & \textquotedbl \\
7 & Ms0.83Ma1.74  & $0.1$ & $0.7$    & $0.83$ & $1.74$ & Transonic and super-\alf ic \\
8 & Ms0.45Ma1.72  & $0.1$ & $2$       & $0.45$ & $1.72$ & Subsonic and super-\alf ic \\
\hline
9   & Ms9.92Ma0.5    & $1$ & $0.0049$ & $9.92$ & $0.50$ & Supersonic and sub-\alf ic \\
10 & Ms7.89Ma0.5    & $1$ & $0.0077$ & $7.89$ & $0.50$ & \textquotedbl \\
11 & Ms6.78Ma0.52  & $1$ & $0.01$   & $6.78$ & $0.52$ & \textquotedbl \\
12 & Ms4.46Ma0.55  & $1$ & $0.025$ & $4.46$ & $0.55$ & \textquotedbl \\
13 & Ms3.16Ma0.58  & $1$ & $0.05$   & $3.16$ & $0.58$ & \textquotedbl \\
14 & Ms2.41Ma0.67  & $1$ & $0.1$     & $2.41$ & $0.67$ & \textquotedbl \\
15 & Ms0.87Ma0.7    & $1$ & $0.7$     & $0.87$ & $0.70$ & Transonic and sub-\alf ic \\
16 & Ms0.48Ma0.65  & $1$ & $2$       & $0.48$ & $0.65$ & Subsonic and sub-\alf ic \\
\hline
17 & Ms8.42Ma0.22  & $3$ & $0.01$  & $8.42$ & $0.22$ & Supersonic and sub-\alf ic \\
18 & Ms8.39Ma0.14  & $5$ & $0.01$  & $8.39$ & $0.14$ & \textquotedbl \\
\hline
\end{tabular}
\end{table*}

Each simulation was allowed to evolve for five eddy turnover times so that the turbulent energy cascade formed over the inertial range, meaning that the turbulence had sufficiently developed. The three-dimensional cubes of thermal electron density, and the x, y and z components of the velocity and magnetic field, were then recorded for the specific instance of turbulence captured in this snapshot.

From this output, the instantaneous sonic and \alf ic Mach numbers of each simulation are calculated according to Eq. \ref{Mach_nums}. The sound speed is calculated from the isothermal equation of state, $c_s = \sqrt{p/\rho}$, applied to the initial conditions of the simulation, as the sound speed should be independent of time for isothermal turbulence. These Mach numbers gradually change over time, due to the random driving of the turbulence. However, we are only interested in the Mach numbers of the turbulence for the specific instance observed, as our goal is to compare statistics of the observed turbulence to the properties of turbulence for this instance. This means that our method of determining the properties of turbulence relies on the assumption that the statistics of synchrotron intensity instantly adjust as the Mach numbers change. As can be seen from Table \ref{TabSims}, low values of the initial pressure produce supersonic simulations, and high initial pressures produced subsonic simulations. If the initial magnetic field strength is low, then the simulation is super-\alf ic, and otherwise it is sub-\alf ic.

To construct maps of synchrotron emission for each simulation, a three-dimensional cube of $B_{\perp}$ was calculated according to $B_{\perp} = \sqrt{B_x^2 + B_y^2}$ for lines of sight along the z-axis, and $B_{\perp} = \sqrt{B_x^2 + B_z^2}$ for lines of sight along the y-axis. We then assumed a fixed observing frequency, and constant $\gamma$ and $K$ within the emitting region, so that the synchrotron intensity 
\begin{equation}
I \propto \int B_{\perp}^{\gamma} \mathrm{d}L, \label{mock_sync}
\end{equation}
and integrated $B_{\perp}^{\gamma}$ along the line of sight. The resulting map was then normalized by dividing the intensity at each pixel by the number of pixels along the line of sight, so that the produced intensity maps do not depend on the size of the simulation. An example synchrotron map is shown in Figure \ref{fig1}, for the Ms0.48Ma0.65 simulation. This map shows large and small scale filaments aligned with the magnetic field.

In Section \ref{los}, we study how statistics of synchrotron intensity depend on the angle between the mean magnetic field and the line of sight. To perform this study, we constructed synchrotron maps for different lines of sight through the simulation cubes. For a line of sight initially along the z-axis, we rotated the cube about the y-axis, and projected the x and z components of the magnetic field onto a vector perpendicular to both the y-axis and the new line of sight. $B_{\perp}$ was then calculated by $B_{\perp} = \sqrt{[B_x \cos(\theta) + B_z \sin(\theta)]^2 + B_y^2}$, for a rotation around the y-axis by angle $\theta$. For a line of sight initially along the y-axis, the same method was used, but with the roles of the y- and z- axes interchanged.

From the cube thus formed, a sub-cube with $362$ pixels along each side was extracted from the centre of the cube, as this is the largest sub-cube that exists for all lines of sight into the cube. The synchrotron map was then produced as before, by integrating $B_{\perp}^{\gamma}$ along the line of sight and dividing by the number of pixels along the line of sight.

\section{Results}
\label{results}

\subsection{Testing Lazarian and Pogosyan 2012}
\label{testLP12}
In this section we will test the predictions made by LP12, using the three-dimensional MHD simulations listed in Table \ref{TabSims}. In particular, we will focus on the predictions that NCFs of synchrotron emissivity and intensity are insensitive to $\gamma$, and that the quadrupole ratio provides a good measure of anisotropy. 

\subsubsection{Correlation Function and Structure Function}
\label{cfClaim}
One of the predictions made by LP12 is that the NCF of synchrotron emissivity, $\xi_{B_{\perp}^{\gamma}}$ can be approximated by Eq. \ref{sync_emis_approx} for isotropic Gaussian turbulence, and any value of $\gamma$ between $1$ and $4$. We first test if this equation is valid for the MHD simulations being considered, by calculating the left and right hand sides of Eq. \ref{sync_emis_approx}. Due to the anisotropy introduced into our simulations by the magnetic field, and the fact that supersonic turbulence is non-Gaussian, we expect that Eq. \ref{sync_emis_approx} should only be approximately true for subsonic, super-\alf ic simulations. For other regimes of turbulence, we do not expect there to be a good match between the left and right hand sides of Eq. \ref{sync_emis_approx}. The left hand side, $\xi_{B_{\perp}^{\gamma}}$ is calculated according to Eq. \ref{norm_corr_emiss}, and the right hand side is calculated using equations \ref{c_1} and \ref{c_2}. {Both the left and right hand sides are then radially averaged, so that they can be plotted as a function of radial separation $r = |\mathbf{r}|$. We assume that $\gamma = 2$ for these calculations, so that there should be equality in Eq. \ref{sync_emis_approx} for isotropic Gaussian turbulence.

These calculations were performed for all of the simulations in Table \ref{TabSims}. In Figure \ref{fig2}, we show representative plots of the left and right hand sides of Eq. \ref{sync_emis_approx} for low magnetic field simulations (top row), and high magnetic field simulations (bottom row). Supersonic simulations are in the left column, and subsonic simulations on the right. We find that for all simulations, there is a discrepancy between the left and right hand sides of Eq. \ref{sync_emis_approx}, that is largest on radial scales between $10$ and $100$ pixels. This discrepancy tends to increase as the sonic Mach number of the simulation increases. Additionally, by studying the corresponding plots for the Ms8.42Ma0.22 and Ms8.39Ma0.14 simulations, we find that this discrepancy tends to slightly increase as the strength of the mean magnetic field increases.

\begin{figure*}
\centering
\includegraphics[scale=0.7, trim=10 0 0 0, clip]{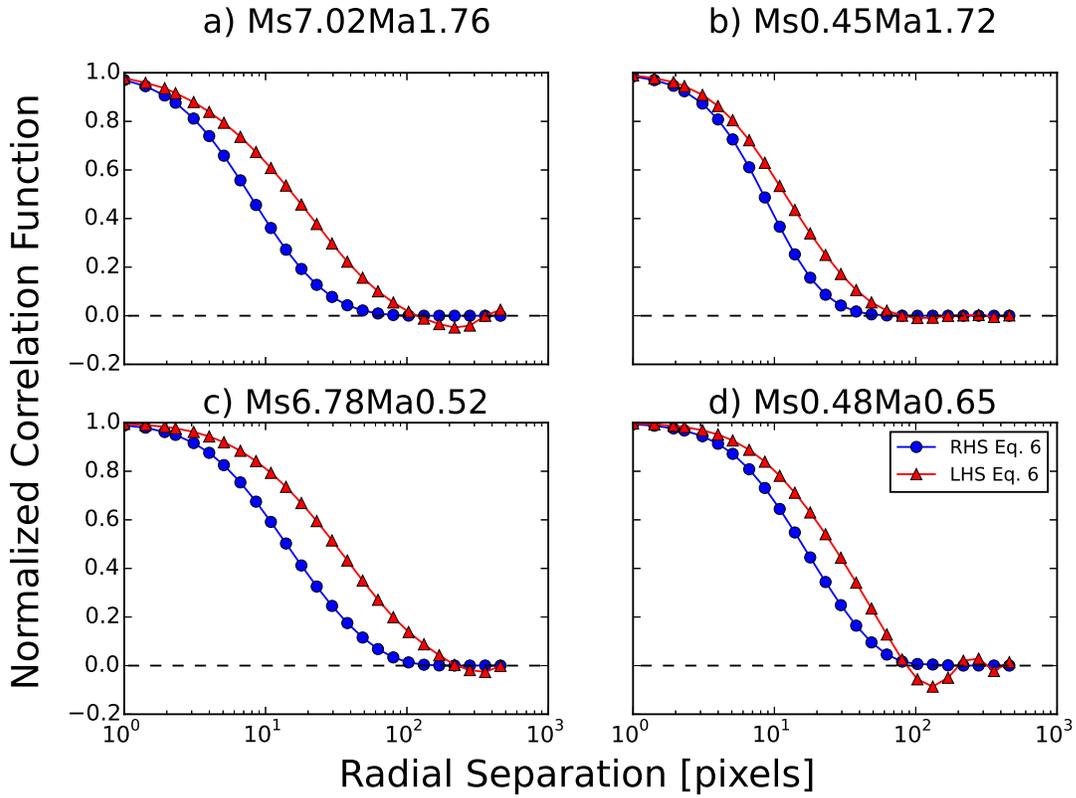}
\caption{Plots of the left and right hand sides of Eq. \ref{sync_emis_approx} for a) simulation 3, Ms7.02Ma1.76, b) simulation 8, Ms0.45Ma1.72, c) simulation 11, Ms6.78Ma0.52, and d) simulation 16, Ms0.48Ma0.65.} \label{fig2}
\end{figure*}

Thus, we find as expected that Eq. \ref{sync_emis_approx} is not applicable to simulations of ideal MHD turbulence. This is because the derivation of Eq. \ref{sync_emis_approx} relies on the assumption that the synchrotron emissivity has an isotropic Gaussian distribution. For supersonic simulations, this assumption is violated because the formation of shocks compresses and amplifies the magnetic field, causing the synchrotron emissivity to deviate from a Gaussian distribution. For sub-\alf ic simulations, the strong magnetic fields produce anisotropic structures in the synchrotron emissivity. Simulations that are subsonic and super-\alf ic also exhibit synchrotron emissivity distributions that deviate from a Gaussian, due to the natural evolution of the turbulence, and this produces a small discrepancy between the left and right hand sides of Eq. \ref{sync_emis_approx}. 

Although the NCF of synchrotron emissivity cannot be calculated via Eq. \ref{sync_emis_approx} for our simulations, it is still possible that the NCF is insensitive to $\gamma$ for $\gamma$ values between $1$ and $4$, as predicted by LP12. To test this prediction, we again calculated the NCF of synchrotron emissivity according to Eq. \ref{norm_corr_emiss}, this time for values of $\gamma$ between $1$ and $4$, in increments of $0.5$. 

\begin{figure*}
\centering
\includegraphics[scale=0.7]{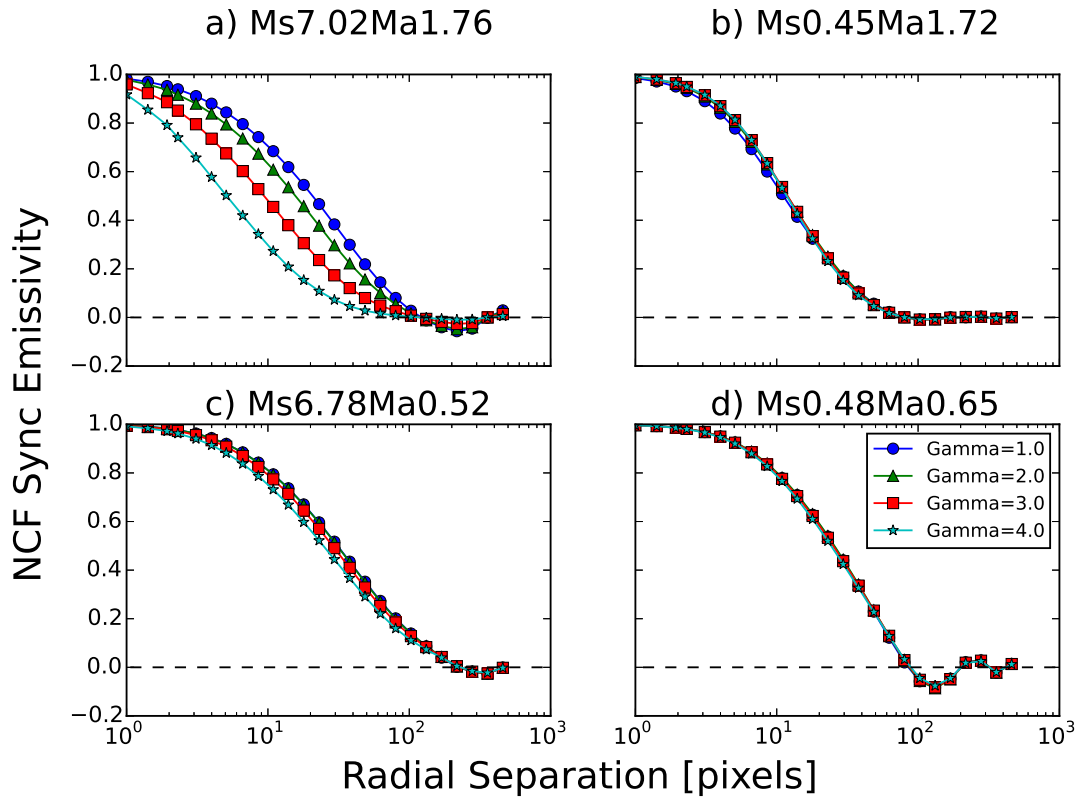}
\caption{The NCF of synchrotron emissivity for the same simulations as in Figure \ref{fig2}. For each simulation, NCFs are produced for $\gamma = 1.0, 2.0, 3.0,$ and $4.0$, colored blue, green, red and cyan respectively.} \label{fig3}
\end{figure*}

In Figure \ref{fig3} we display the NCFs of synchrotron emissivity for $\gamma$ values of $1, 2, 3$ and $4$, for four simulations. Simulations with a weak magnetic field are in the top row, and simulations with a strong magnetic field are in the bottom row. The left column displays the NCFs for supersonic simulations, and the right column for subsonic simulations. As shown in Figure \ref{fig3}a), we find that the NCF of synchrotron emissivity exhibits significant $\gamma$ dependence for simulations that are supersonic and have a weak magnetic field. By examining the NCFs for the eight simulations with an initial magnetic field strength of $0.1$, we also find that the degree of $\gamma$ dependence decreases as the sonic Mach number decreases. In particular, there is significantly reduced $\gamma$ dependence for subsonic simulations, as seen in Figure \ref{fig3}b). 

For simulations with an initial magnetic field strength of $1$, we find that the degree of $\gamma$ dependence is always less than for the corresponding simulation with the same initial pressure, but initial magnetic field strength of $0.1$. This is most clearly seen in Figure \ref{fig3}c). As for the low magnetic field simulations, we find that there is less $\gamma$ dependence as the sonic Mach number decreases. In addition, we find that there is no visible $\gamma$ dependence for the Ms8.42Ma0.22 and Ms8.39Ma0.14 simulations.

Thus, we find that for simulations with a significantly strong magnetic field, or high pressure, the assumption that the NCF of synchrotron emissivity is independent of $\gamma$ is valid. However, for supersonic simulations involving a weak magnetic field, the NCF of synchrotron emissivity is sensitive to $\gamma$. This result differs from the theoretical findings of LP12 because the formation of shocks causes the distribution of synchrotron emissivity to become non-Gaussian, violating the assumption of an isotropic Gaussian distribution. 

LP12 also propose that the NCFs of synchrotron intensity are insensitive to $\gamma$. As the synchrotron intensity is an observable quantity, this is the prediction most relevant to this paper, out of the five made by LP12. To test this prediction, we integrated $B_{\perp}^{\gamma}$ (i.e. the synchrotron emissivity, see equation \ref{mock_sync}) along the line of sight for all simulations, for $\gamma$ values between $1$ and $4$, in increments of $0.5$. We only analysed synchrotron maps produced for a line of sight along the z-axis, to enable a direct comparison to the predictions made by LP12.

In Figure \ref{fig4} we display the NCFs of synchrotron intensity for the same simulations as in Figure \ref{fig3}, for $\gamma$ values of $1, 2, 3$ and $4$. Similar to the results found for synchrotron emissivity, we find that for simulations with an initial magnetic field strength of $0.1$, the degree of $\gamma$ dependence decreases as the sonic Mach number decreases. We believe that this $\gamma$ dependence arises for the same reasons as discussed for the NCFs of synchrotron emissivity, namely that the shocks generated in supersonic turbulence create a non-Gaussian distribution of synchrotron intensity. We also find that the Ms8.42Ma0.22 and Ms8.39Ma0.14 simulations have very little dependence on $\gamma$, as was found for synchrotron emissivity. However, unlike the results found for the NCF of synchrotron emissivity, the NCFs of synchrotron intensity for simulations with an initial magnetic field strength of $1$ do not display any clear $\gamma$ dependence in general, except for the Ms0.87Ma0.7 and Ms0.48Ma0.65 simulations. These simulations are transonic and subsonic respectively, and so we conclude that the strong magnetic fields of these simulations are causing anisotropic structures to survive the turbulent motion of the plasma. These structures are likely the cause of the weak $\gamma$ dependence observed for the NCFs of the synchrotron intensity maps of the Ms0.87Ma0.7 and Ms0.48Ma0.65 simulations. We postulate that the reason why the NCFs for synchrotron intensity of these simulations show some mild $\gamma$ dependence, but NCFs of synchrotron emissivity do not, is that integrating the synchrotron emissivity along the line of sight makes the anisotropy easier to detect with the correlation function, by superimposing cross-sections of synchrotron emitting filaments, where each cross-section is anisotropic. 

Thus, we find that the NCF of synchrotron intensity is insensitive to $\gamma$ for most regimes of turbulence. If the turbulence is supersonic and has a weak mean magnetic field, then the NCF of synchrotron intensity does have some dependence on $\gamma$. However, if we only consider values of $\gamma$ that are likely to be observed, between $1.5$ and $2.5$, say, then the NCF of synchrotron intensity does not exhibit significant dependence on $\gamma$ for any regime of turbulence. Hence, we can compare simulated and observed NCFs of synchrotron intensity without needing to take $\gamma$ into account, in most situations. 

\begin{figure*}
\centering
\includegraphics[scale=0.7]{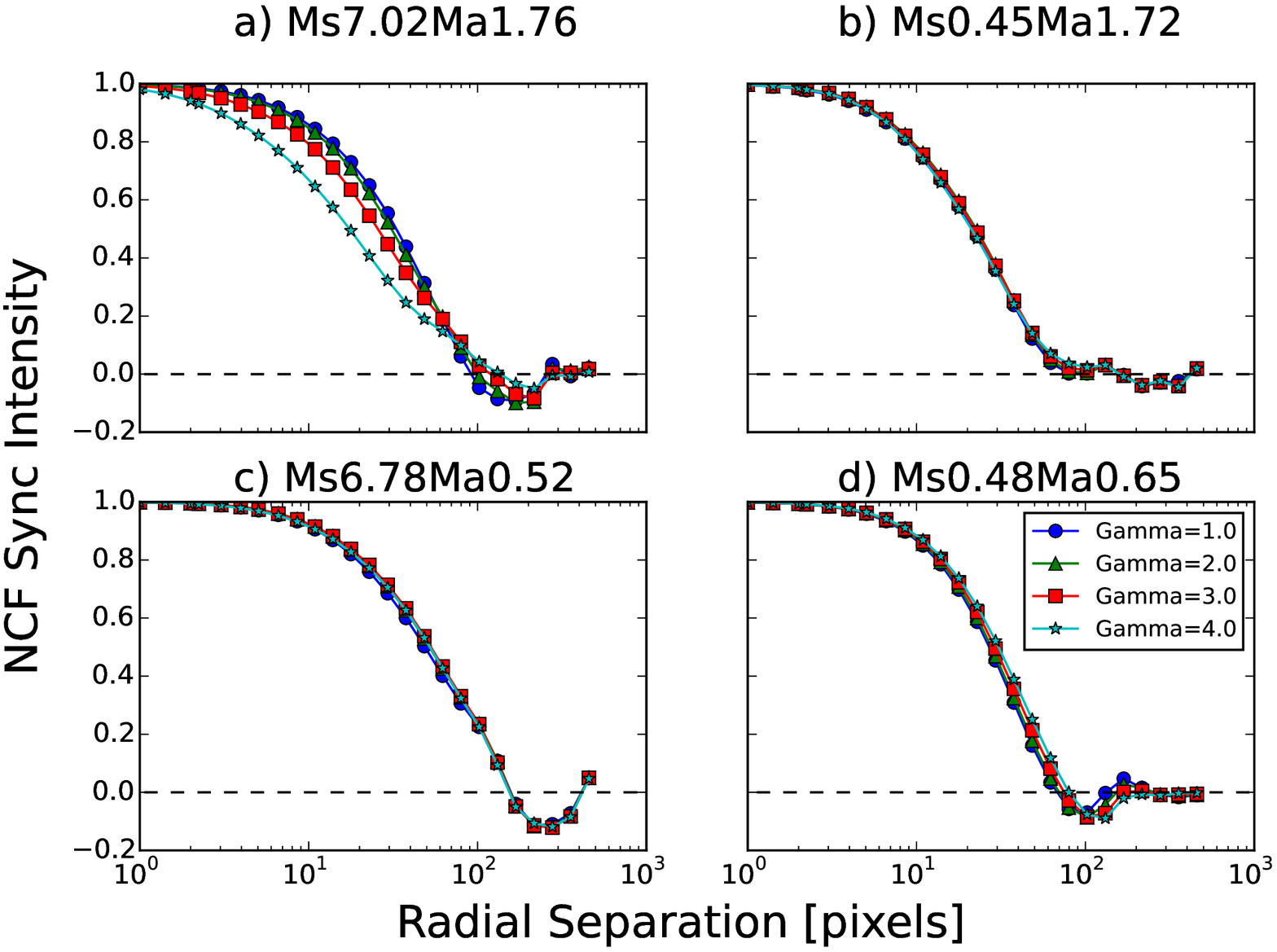}
\caption{The NCF of synchrotron intensity for the same simulations as in Figure \ref{fig2}. For each simulation, NCFs are produced for $\gamma = 1.0, 2.0, 3.0,$ and $4.0$, colored blue, green, red and cyan respectively.} \label{fig4}
\end{figure*}

\subsubsection{Quadrupole Ratio}
\label{quadClaim}
To quantify the degree of anisotropy present in a map of synchrotron intensity, LP12 introduced the quadrupole ratio. Additionally, LP12 estimated that the quadrupole ratio modulus should be less than $0.45$, approximately $0.4$ for sub-\alf ic turbulence, $0.3$ for trans-\alf ic turbulence, and $0.1$ for super-\alf ic turbulence. In this section we will test these predictions with the simulations of ideal MHD turbulence listed in Table \ref{TabSims}, and assess the viability of the quadrupole ratio as a descriptor of anisotropy.

We calculate the quadrupole ratio according to Eq. \ref{quad_ratio}, by first calculating the normalized structure function of synchrotron intensity using Eq. \ref{norm_SF}. These calculations are performed for $\gamma$ values between $1$ and $4$, in increments of $0.5$.  

\begin{figure*}
\centering
\includegraphics[scale=0.7]{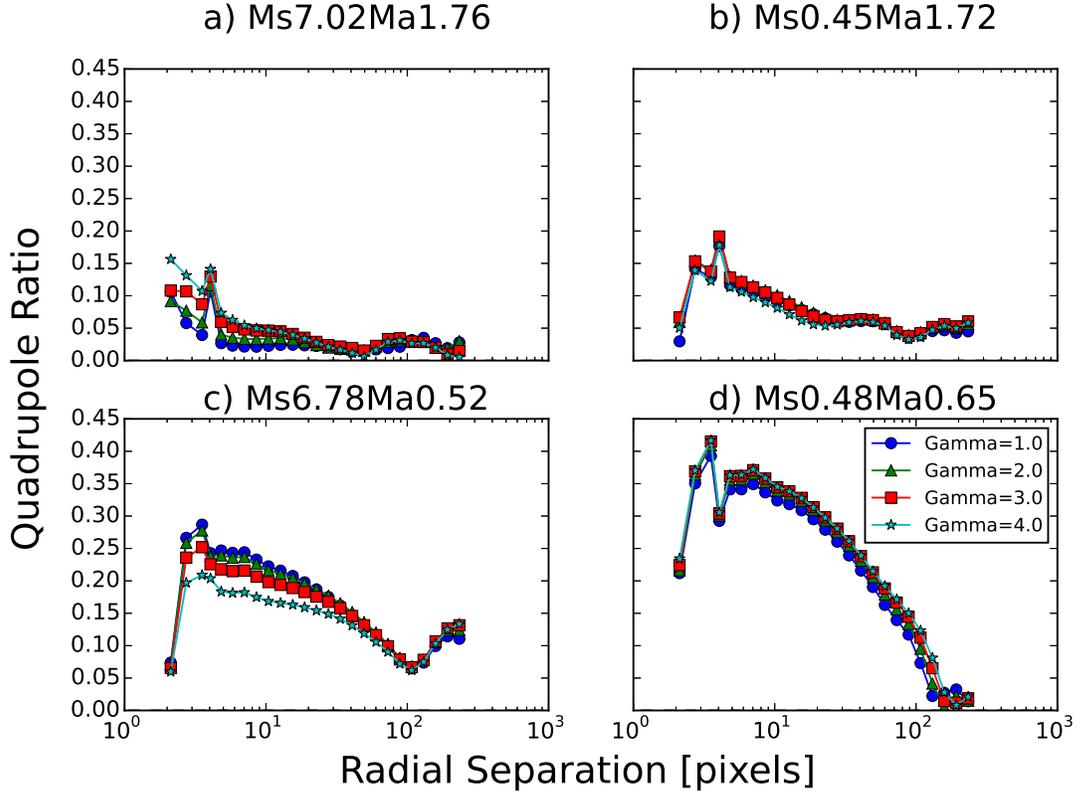}
\caption{The quadrupole ratio modulus of synchrotron intensity for the same simulations as in Figure \ref{fig2}. For each simulation, quadrupole ratios are produced for $\gamma = 1.0, 2.0, 3.0,$ and $4.0$, colored blue, green, red and cyan respectively.} \label{fig5}
\end{figure*}

In Figure \ref{fig5} we show the quadrupole ratio moduli calculated for the same simulations as in Figure \ref{fig3}, for the same values of $\gamma$. We find that all quadrupole ratios depend on the radial separation of the points used to calculate the structure function, unlike the results found by LP12. In particular, the quadrupole ratio modulus tends to decrease on larger radial scales, indicating that there is less anisotropy on these scales. This could be caused by how the MHD simulations are performed; the energy injection scale for the simulations is approximately $200$ pixels (this can differ between simulations), and there is no anisotropy on this scale, because the turbulence has not fully developed. The anisotropy takes time to form down the cascade, which might be why the degree of anisotropy generally decreases with increasing radial separation.

Many of the plots of the quadrupole ratio modulus also display a prominent peak or trough at a radial separation of approximately $4$ pixels. These features arise because the two-dimensional structure function exists on a discrete grid of pixels, and the quadrupole ratio is calculated from this by integrating over pixels that fall within a thin annulus. The pixels of the structure function being used to calculate the quadrupole ratio for a radial separation of $4$ pixels all lie close to the diagonals of the structure function, as no other pixels fall into this radial bin. The imaginary part of the quadrupole ratio is particularly sensitive to pixels along the diagonals, due to its $\sin(2\phi)$ dependence (Eq. \ref{quad_ratio}). Hence, the real part of the quadrupole ratio (which should be the larger component based on the set-up of our simulation) is not well probed at this radial separation, and the measured quadrupole ratio is overly sensitive to the small imaginary part, which traces diagonal anisotropy. This anisotropy only exists because of random fluctuations in the turbulence, arbitrarily leading to peaks or troughs in the quadrupole ratio modulus.

As illustrated in Figure \ref{fig5}, we find that for a fixed initial pressure, simulations with a higher magnetic field strength (lower \alf ic Mach number) have a larger quadrupole ratio modulus. This is because stronger magnetic fields produce more anisotropic structures, indicating that the quadrupole ratio is a reliable measure of anisotropy. We also find that for a fixed initial magnetic field strength, simulations with a lower sonic Mach number (higher pressure) have a larger quadrupole ratio modulus, and hence exhibit higher anisotropy. This is because the plasma generally has lower velocity in these simulations, shocks do not form, and so the magnetic field is more capable of constraining the flow of plasma to produce anisotropic structures. 

The quadrupole ratios calculated for the majority of the simulations also exhibit some slight $\gamma$ dependence, however there does not seem to be any clear relationship between the sonic and \alf ic Mach numbers of the turbulence and how $\gamma$ affects the quadrupole ratio. How $\gamma$ affects the measured quadrupole ratio may depend upon the particular instance of turbulence observed, and so $\gamma$ must be measured in order to compare observations to simulations. The two simulations that do not show any $\gamma$ dependence are the Ms8.42Ma0.22 and Ms8.39Ma0.14 simulations, as was found for the NCF of synchrotron intensity. 

Out of all the simulations studied, the Ms8.39Ma0.14 simulation has the greatest quadrupole ratio modulus (not shown here), which has an amplitude of approximately $0.37$ up to a radial separation of approximately $100$ pixels. Thus, we support the prediction that the quadrupole ratio modulus should be less than $0.45$ for all simulations, and the prediction that sub-\alf ic simulations have quadrupole ratio moduli that are approximately $0.4$. Our only supersonic, trans-\alf ic simulation is the Ms10.96Ma1.41 simulation, and this simulation has a quadrupole ratio modulus of $0.15$ or less. This is much smaller than the value estimated by LP12 of $0.3$, however this difference is likely due to the Ms10.96Ma1.41 simulation being highly supersonic, with a weak initial magnetic field. The simulations that are closest to being subsonic and trans-\alf ic, Ms0.87Ma0.7 and Ms0.48Ma0.65, have quadrupole ratio moduli that are between $0.2$ and $0.35$ for a wide range of radial separations. Hence, we also support the prediction that simulations of trans-\alf ic turbulence should have quadrupole ratio moduli of approximately $0.3$. Finally, our super-\alf ic simulations all exhibit quadrupole ratio moduli between $0$ and $0.15$, and thus our findings support the prediction that the quadrupole ratio modulus for simulations of super-\alf ic turbulence is approximately $0.1$.

\subsection{Statistical Diagnostics of Parameters of Turbulence}
\label{statDiag}
In this section we will compare statistics of the mock synchrotron intensity maps to the sonic and \alf ic Mach numbers of the corresponding simulations, to investigate whether these properties of turbulence can be derived from synchrotron intensity maps. The statistics we examine include the structure function slope (via the variable $m$, the structure function slope minus $1$), and the integrated quadrupole ratio modulus of synchrotron intensity.  
However, as shown in Section \ref{testLP12}, the NCF of synchrotron intensity has a weak dependence on $\gamma$, and thus it is possible that the statistics we study in this section also depend on $\gamma$, as they are both related to the NCF. In Appendix \ref{GamDep} we show that $m$ and the integrated quadrupole ratio modulus depend only weakly on $\gamma$, provided the regime of turbulence is not supersonic and super-\alf ic, and thus the value of $\gamma$ in an observed region of sky should not interfere with any procedure designed to determine the sonic and \alf ic Mach numbers from these statistics in general. Henceforth, we consider the mock synchrotron maps produced for $\gamma = 2$, as this value is typical of the Milky Way and often assumed in analytical work, and compare the statistics of these maps to the sonic and \alf ic Mach numbers. We also examine the skewness and kurtosis of synchrotron intensity in Appendix \ref{AppSkew}, and find that they are not sensitive to the sonic or \alf ic Mach numbers.

For each simulation, each statistic is calculated for the synchrotron maps produced for lines of sight along the y- and z-axes, and the average of these is plotted in the following figures. To provide uncertainties on each of these statistics, we divided the synchrotron maps obtained for lines of sight along the y- and z-axes into quarters, and calculated the statistics for each of the eight quarters obtained. The standard error of the mean of the statistics calculated for the quarters was then used to define error bars on the following figures. To test the reliability of these statistics, in Appendix \ref{AppSimSize} we examine how sensitive the statistics are to changes in the size of the simulation cube used to produce synchrotron intensity maps. We find that for the integrated quadrupole ratio modulus, all but two simulations appear to be converging to a final value, and so we believe that these statistics are reliable. However, there are significant changes in the structure function slope for small changes in the size of the simulation cube for high magnetic field simulations, and thus the values of $m$ shown in the following figures should be regarded as tentative. We find that the error bars for the structure function slope and integrated quadrupole ratio modulus are consistent with the variation seen in these statistics as the size of the simulation cube changes, and we hence believe that the error bars provide an adequate depiction of the reliability of the data.

\subsubsection{Structure Function Slope}
\label{sfslope}
We calculate the structure function of our mock synchrotron intensity maps according to Eq. \ref{sync_SF} for all simulations (examples can be found in Figure \ref{fig11}). The slopes of these structure functions are measured by performing a linear fit to the structure function for radial separations between 20 and 50 pixels, as this range corresponds to the inertial range of the simulations, whilst avoiding the radial scales where the structure functions begin to plateau. We express the structure function slope via the variable $m$, which is the structure function slope minus $1$. 

In Figure \ref{fig6} we plot $m$ against the sonic Mach number (left), and the \alf ic Mach number (right). Simulations with $B=0.1$ are plotted as blue circles, simulations with $B=1$ as red stars, and the remaining simulations as green triangles. We find that the simulations fall into two distinct populations in Figure \ref{fig6}a; with high magnetic field simulations having a larger value of $m$ than low magnetic field simulations. For both high ($B=1$, red) and low ($B=0.1$, blue) magnetic field simulations, there does not appear to be any dependence of $m$ on the sonic Mach number. 

\begin{figure}
\centering
\includegraphics[scale=0.44, trim=28 0 0 0, clip]{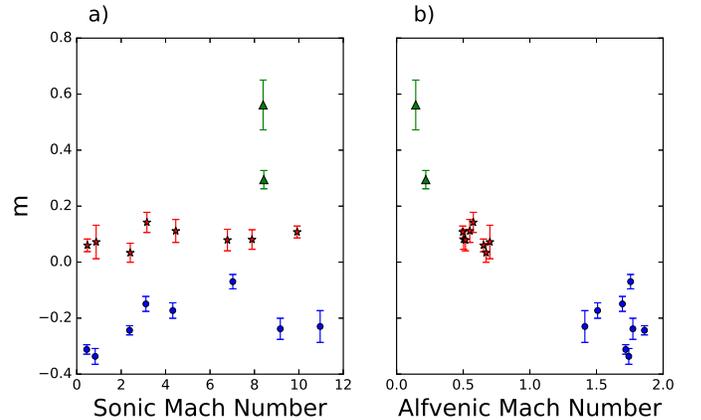}
\caption{The structure function slope minus $1$ ($m$) of synchrotron intensity against a) sonic Mach number and b) \alf ic Mach number for all simulations, and $\gamma = 2$. Simulations with $B=0.1$ are plotted as blue circles, simulations with $B=1$ as red stars, and the remaining simulations as green triangles.} \label{fig6}
\end{figure}

We do find a linear trend between $m$ and the \alf ic Mach number, with some scatter about this trend due to the different sonic Mach numbers of the simulations. This linear correlation between $m$ and the \alf ic Mach number is also found for $\gamma = 1$, and so we conclude that this result is not caused by any dependence of $m$ on $\gamma$. Hence, we find that the structure function slope is sensitive to the \alf ic Mach number, and so should be a useful statistic for determining this parameter of turbulence.

\subsubsection{Integrated Quadrupole Ratio}
\label{intquad}
We calculate the quadrupole ratio for each simulation using Eq. \ref{quad_ratio}. To reduce the quadrupole ratio modulus plots to a single number that can be compared to the sonic and \alf ic Mach numbers, we integrated the quadrupole ratio modulus over the inertial range of the turbulence. For our simulations, this means we integrated from $\sim 20$ pixels, above the dissipation scale of the turbulence, to $\sim 100$ pixels, below the energy injection scale. This integration was performed over the logarithm of the radial separation, so that we calculated the area under the plots of the quadrupole ratio modulus as they appear in Figure \ref{fig5}. We then normalized this area by dividing by the number of points used in the calculation of the integral, so that we essentially calculate the average of the quadrupole ratio modulus over this radial separation range. We call this quantity the integrated quadrupole ratio modulus.

\begin{figure}
\centering
\includegraphics[scale=0.42, trim=25 0 0 0, clip]{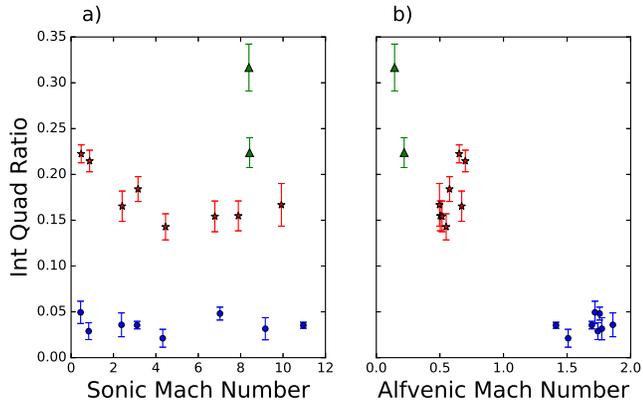}
\caption{The integrated quadrupole ratio modulus of synchrotron intensity against a) sonic Mach number and b) \alf ic Mach number for all simulations, and $\gamma = 2$. Simulations with $B=0.1$ are plotted as blue circles, simulations with $B=1$ as red stars, and the remaining simulations as green triangles.} \label{fig7}
\end{figure}

In Figure \ref{fig7} we plot the integrated quadrupole ratio modulus against the sonic Mach number (left), and \alf ic Mach number (right), similar to Figure \ref{fig6}. As was the case for the structure function slope, we find that the simulations separate into two distinct groups when we plot the integrated quadrupole ratio modulus against sonic Mach number. The super-\alf ic simulations (blue) have small integrated quadrupole ratio moduli for all sonic Mach numbers studied, with no discernible dependence on the value of the sonic Mach number. This is because if the magnetic field is too weak to cause anisotropy in the observed synchrotron intensity structures, then changing the sonic Mach number of the turbulence, i.e. the fluid velocity, will not introduce any anisotropy. The sub-\alf ic simulations (red and green) have larger integrated quadrupole ratio moduli than the super-\alf ic simulations, because the large initial magnetic field strength produces anisotropic structures. There is also some evidence of the integrated quadrupole ratio modulus depending on the sonic Mach number for sub-\alf ic simulations (red data points). This dependence is clearest for small \alf ic Mach numbers, where the integrated quadrupole ratio modulus decreases with increasing sonic Mach number. This occurs because increasing the sonic Mach number corresponds to faster fluid motions, and these motions can tear apart and redirect filaments of ionized gas, thus decreasing the observed anisotropy. However, the integrated quadrupole ratio modulus only decreases with sonic Mach number up to a sonic Mach number of about $4$; above this sonic Mach number there is no clear correlation between these variables. This could indicate that there is a limit to how much anisotropy can be erased by supersonic fluid motions. We also find that there is a linear correlation between the integrated quadrupole ratio modulus and the \alf ic Mach number, with the scatter likely being caused by the different sonic Mach numbers of the simulations. In summary, we find that the integrated quadrupole ratio modulus is sensitive to the \alf ic Mach number, with some mild dependence on the sonic Mach number, and so it is a useful statistic to use when trying to characterize observed turbulence.

\subsection{Dependence of Statistics on Line of Sight}
\label{los}
The results shown so far have all been produced for lines of sight that are perpendicular to the mean magnetic field of the simulation. This represents the optimal scenario for observing the effect of the magnetic field, as we are best able to see the anisotropy that is introduced by the magnetic field. However, for a line of sight that is parallel to the mean magnetic field, we should not observe any clear anisotropy, and thus the statistics we have examined should depend upon the orientation of the mean magnetic field relative to our line of sight. In this section we examine the dependence of the statistics studied so far on the relative orientation of the mean magnetic field. We perform this study by creating mock synchrotron intensity maps for different lines of sight into the simulation cubes, as described in Section \ref{method}, using $\gamma = 2$. We then calculate the structure function slope and integrated quadrupole ratio modulus for each map produced (skewness and kurtosis are examined in Appendix \ref{AppSkew}). If a statistic is particularly sensitive to the relative orientation of the mean magnetic field, then this could allow us to use this statistic to determine what the relative orientation is. We also study whether these statistics maintain sensitivity to the sonic and \alf ic Mach numbers of the turbulence for different lines of sight. Because the statistics we calculate in this section are calculated from sub-cubes, not the full simulation cubes, the value of the statistics will differ from those calculated in Section \ref{statDiag}. As the results of Section \ref{statDiag} were calculated using more data, they are more reliable, and so take precedence whenever there is a conflict between the results of this section and prior results.

\subsubsection{Structure Function Slope}
\label{sfslope_los}
In Figure \ref{fig8} we plot $m$ against the relative angle between the line of sight and the mean magnetic field, for low magnetic field simulations on the left, and high magnetic field simulations on the right. A relative angle of $0^{\circ}$ means that the line of sight is parallel to the mean magnetic field, and an angle of $90^{\circ}$ means the line of sight is perpendicular to the mean magnetic field. We find that $m$ does not show any clear dependence on the relative orientation of the mean magnetic field for low magnetic field simulations. We also find that the structure function slope tends to increase as the line of sight becomes more perpendicular to the mean magnetic field for high magnetic field simulations. This result is expected, as Figure \ref{fig6} demonstrates that simulations with a strong magnetic field perpendicular to the line of sight, which are sub-\alf ic, should have larger values of $m$ than simulations with a weak magnetic field perpendicular to the line of sight. As the component of the magnetic field perpendicular to our line of sight increases as the line of sight becomes more perpendicular to the mean magnetic field, we expect to see $m$ increase as well. Finally, we find that $m$ loses sensitivity to the \alf ic Mach number as the line of sight becomes parallel to the mean magnetic field. Hence, the structure function slope of synchrotron intensity is only a useful statistic if the magnetic field strength perpendicular to our line of sight is sufficiently strong.

\begin{figure}
\centering
\includegraphics[scale=0.42, trim=30 0 0 0, clip]{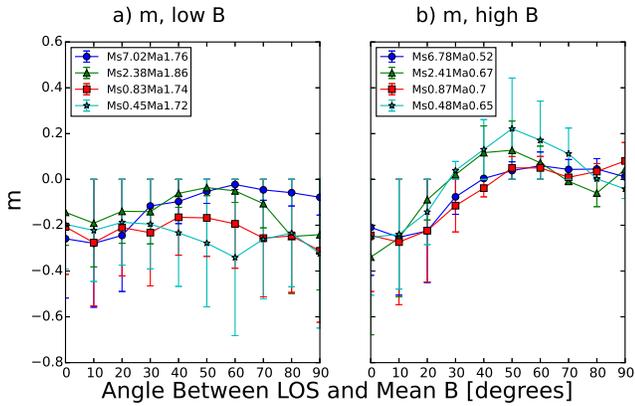}
\caption{The structure function slope minus $1$ ($m$) of synchrotron intensity against the relative angle between the line of sight and the mean magnetic field. The low magnetic field simulations used are Ms7.02Ma1.76 (blue), Ms2.38Ma1.86 (green), Ms0.83Ma1.74 (red) and Ms0.45Ma1.72 (cyan). The high magnetic field simulations used are Ms6.78Ma0.52 (blue), Ms2.41Ma0.67 (green), Ms0.87Ma0.7 (red) and Ms0.48Ms0.65 (cyan). a) Low B and b) high B simulations.} \label{fig8}
\end{figure}

\subsubsection{Integrated Quadrupole Ratio}
\label{intquad_los}
In Figure \ref{fig9} we plot the integrated quadrupole ratio modulus against the relative angle between the line of sight and the mean magnetic field, for low magnetic field simulations on the left, and for high magnetic field simulations on the right. The same simulations are used as in Figure \ref{fig8}. As was found for $m$, the integrated quadrupole ratio modulus does not show any clear dependence on the orientation of the mean magnetic field relative to the line of sight for low magnetic field simulations. This is because these simulations do not exhibit much anisotropy due to their weak magnetic fields, and so the statistics calculated should be approximately the same for all lines of sight. The value of the integrated quadrupole ratio modulus remains non-zero even for lines of sight parallel to the mean magnetic field because of statistical effects. We also find that for simulations with a strong mean magnetic field, the integrated quadrupole ratio modulus increases as the line of sight becomes more perpendicular to the mean magnetic field. This occurs because more anisotropy is visible as the line of sight becomes more perpendicular to the mean magnetic field, as expected.

\begin{figure}
\centering
\includegraphics[scale=0.42, trim=25 0 0 0, clip]{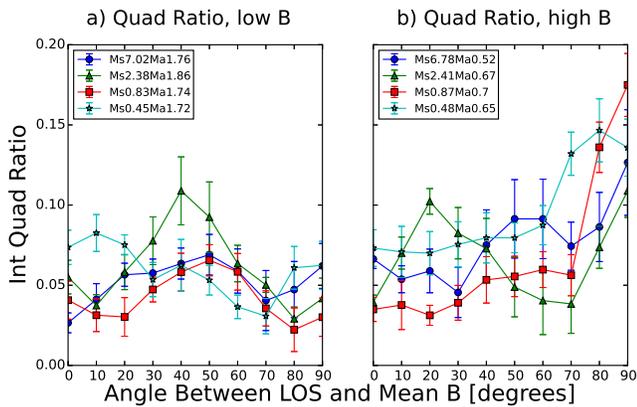}
\caption{Same as Fig. \ref{fig8}, but plotting the integrated quadrupole ratio against the relative angle between the line of sight and the mean magnetic field.} \label{fig9}
\end{figure}

We observe that the integrated quadrupole ratio modulus does not exhibit any clear dependence on the sonic Mach number for any of the lines of sight studied, and that it loses all dependence on the \alf ic Mach number when the line of sight is parallel to the mean magnetic field. Thus, we conclude that the integrated quadrupole ratio modulus can only be used to constrain the \alf ic Mach number if the magnetic field perpendicular to the line of sight is sufficiently strong. Based on the results of Section \ref{intquad}, it can only be used to constrain the sonic Mach number if the component of the magnetic field perpendicular to the line of sight is very strong, and the turbulence is sub- to transonic.

\section{Discussion}
\label{discuss}

As shown in Section \ref{testLP12}, the NCF of synchrotron intensity is approximately independent of $\gamma$ for $\gamma$ values typical of the Milky Way, between $1.5$ and $2.5$, and for all regimes of turbulence. This means that the NCF of synchrotron intensity is a useful statistic for comparing analytical, numerical and observational studies of synchrotron emission, as assumptions (such as assuming that $\gamma = 2$) can be made without influencing the analysis.

Similarly, in Appendix \ref{GamDep} the structure function slope and integrated quadrupole ratio modulus were shown to weakly depend on $\gamma$, in general, for values of $\gamma$ typical of the Milky Way. Thus, statistics measured from observations can be compared to those simulated for $\gamma = 2$, as derived properties of turbulence would only be slightly improved by comparing measured statistics to those simulated for the observed value of $\gamma$. The exception to this is the structure function slope for supersonic, super-\alf ic turbulence, which exhibits significant dependence on $\gamma$. If this statistic is to be applied to an observation of turbulence that may be in this regime, then it is necessary to measure the value of $\gamma$, so that the observed statistics can be compared to simulated statistics for the same value of $\gamma$.

In Appendix \ref{AppSkew} we find that the skewness and kurtosis of synchrotron intensity have no clear dependence on sonic Mach number, and only a weak dependence on the \alf ic Mach number. As these statistics are not very sensitive to these Mach numbers, they are not useful statistics for our purposes. This illustrates that the skewness and kurtosis of synchrotron intensity behave differently to the skewness and kurtosis of the observed polarisation gradient, which were shown by \cite{Burkhart2012} to be sensitive to the sonic Mach number. This difference arises because the total synchrotron intensity is only sensitive to the magnetic field perpendicular to the line of sight, whereas the polarisation gradient is largely caused by rotation of the plane of polarisation, which traces the electron density and the magnetic field parallel to our line of sight.

The structure function slope of synchrotron intensity, however, is sensitive to the \alf ic Mach number, and the integrated quadrupole ratio modulus is also sensitive to the \alf ic Mach number, with some dependence on the sonic Mach number. These two statistics present the greatest potential for determining the properties of turbulence from observations of synchrotron fluctuations. However, the dependence of the structure function slope on the \alf ic Mach number should be regarded as tentative, as we found that the structure function slope was sensitive to changes in the simulation size in Appendix \ref{AppSimSize}. We believe that the measured values of the structure function slope are sensitive to changes in the size of the simulation cube because the nonlocality of MHD turbulence can cause the measured value of $m$ to differ from the true value, due to improper formation of the inertial range of the turbulence \citep{Beresnyak2009}. For example, \cite{Kritsuk2007} found that simulations with $2048$ pixels on each side were required to observe the inertial range for compressible hydrodynamic turbulence, and so simulations of at least this size are likely required to observe the inertial range in simulations of MHD turbulence, and hence to measure $m$ reliably.

Additionally, the relative orientation of the mean magnetic field relative to the line of sight is a third variable that needs to be determined. Not only does this relative orientation affect the values of the structure function slope and integrated quadrupole ratio modulus, but it can also remove the sensitivity of these statistics to the Mach numbers. For example, when the line of sight is parallel to the mean magnetic field, the integrated quadrupole ratio modulus has no dependence on either Mach number. Thus, the number of useful statistics that we can obtain from observations of synchrotron intensity can decrease if the line of sight is parallel to the mean magnetic field, making it more difficult to determine the properties of the turbulence. This occurs because there is a degeneracy between a weak magnetic field perpendicular to the line of sight, and a strong magnetic field parallel to the line of sight. As a result of this degeneracy, it is only possible to use the synchrotron intensity statistics discussed in this paper to determine the properties of an observed turbulent region if we know a priori that the magnetic field in that region is not parallel to our line of sight. This could be confirmed by observing polarization rotation measures towards pulsars or extragalactic sources, the Zeeman splitting of spectral lines, starlight polarized by intervening dust grains, or polarized thermal dust emission.

If the orientation of the magnetic field is not known, then we must study other properties of the observed synchrotron emission, such as its polarization, to break this degeneracy, and determine the properties of the observed turbulence. As \cite{Burkhart2012} have shown that the skewness and kurtosis of the observed polarization gradient are related to the sonic Mach number of the turbulence, we believe that analyzing the statistics of diffuse polarized synchrotron emission will enable us to determine the properties of an observed turbulent, synchrotron emitting region. In a future paper we will extend the work of \cite{Burkhart2012} by calculating the skewness, kurtosis, structure function slope and integrated quadrupole ratio modulus of the polarization gradient, and other new polarization diagnostics, of diffuse, polarized synchrotron emission. This study will hopefully produce many more useful statistics to break the degeneracies that currently exist.

It is possible that the statistics calculated in this paper depend on properties of the turbulence other than the sonic and \alf ic Mach numbers, for example the compressibility of the plasma, and the ratio of the random to total magnetic field strengths. Although we are not aware of any formula that could be used to calculate the compressibility of the turbulence in our simulations, it is possible to calculate the ratio of the random to total magnetic field. Shown in Figure \ref{fig10} is a plot of the mean ratio of the random to total magnetic field strength as a function of the \alf ic Mach number, for all of the simulations in Table \ref{TabSims}. The strength of the random component of the magnetic field was determined by subtracting the mean magnetic field in each of the Cartesian directions from the magnetic field vector at each pixel, and then calculating the magnitude of the resulting random magnetic field vector. The random magnetic field strength at each pixel was then divided by the strength of the total magnetic field at that pixel, and the average ratio of the random to total magnetic field strength calculated over all pixels.

\begin{figure}
\centering
\includegraphics[scale=0.56]{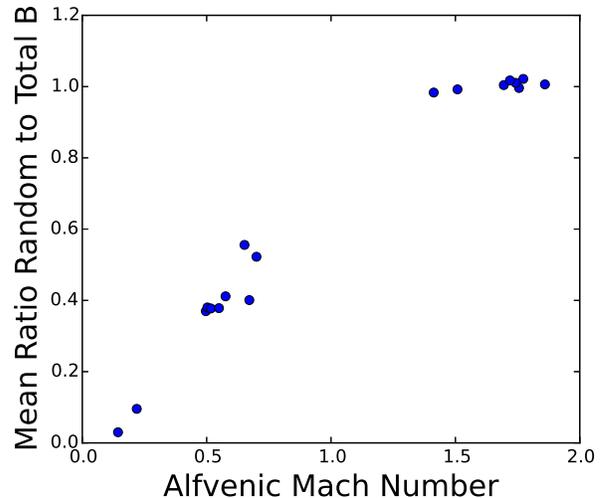}
\caption{The mean ratio of the random magnetic field strength to the total magnetic field strength as a function of the \alf ic Mach number for all simulations.} \label{fig10}
\end{figure}

We find that there is a monotonic relationship between the ratio of the random magnetic field strength to the total magnetic field strength and the \alf ic Mach number. The ratio is small for sub-\alf ic simulations because there is a strong mean magnetic field, and the ratio is close to one for super-\alf ic simulations because there is a small, mostly random magnetic field. Hence, we should be able to deduce the ratio of the random to total magnetic field strength once the \alf ic Mach number is known. It is possible that other parameters of turbulence can also be determined from the sonic or \alf ic Mach numbers.

We still need to consider whether the structure function slope and integrated quadrupole ratio modulus provide meaningful constraints on the sonic and \alf ic Mach numbers, if applied to current observations of diffuse synchrotron intensity. In Appendix \ref{AppObs} we study how observational noise and angular resolution affect the measured structure function slope and integrated quadrupole ratio modulus. We found that a signal-to-noise ratio of $30$, and an angular resolution of less than $7$ pixels were required for the structure function slope and integrated quadrupole ratio modulus to distinguish regimes of turbulence. We require an angular resolution that is capable of resolving the outer-scale of turbulence, and a significant portion of the inertial range of the turbulence, so that the turbulent cascade is resolved spatially. \cite{Haverkorn2008} found that the outer scale of turbulence in the warm ionized medium inside the spiral arms of the Milky Way is on the order of a few pc, say $3$ pc, and so we estimate that a spatial resolution of $0.3$ pc seems capable of resolving synchrotron fluctuations. For an emitting region that is $2$ kpc away, this translates to an angular resolution of $31$ arcseconds. This is in good agreement with our estimate of $21$ arc seconds in Appendix \ref{angres}, based on the requirement that the structure function slope and integrated quadrupole ratio modulus can distinguish between different regimes of turbulence. 

To calculate the sensitivity that is required to observe synchrotron fluctuations, we need to consider resolving local maxima and minima in the synchrotron intensity, as the signal we are interested in is the difference in intensity between these maxima and minima. For a pathlength of $2$ kpc through a turbulent synchrotron emitting medium, where the mean value of $B_{\perp}$ is $3 \, \mu$G, the local maxima of synchrotron intensity observed at a frequency of $1.4$ GHz and with $\gamma = 1.7$ should be $64 \, \mu$Jy/beam, at a resolution of $21$ arc seconds. For fluctuations in synchrotron intensity of $50\%$, we need to resolve fluctuations in intensity of $32 \, \mu$Jy/beam with a signal-to-noise ratio of $30$, and hence the required sensitivity is on the order of $1 \, \mu$Jy/beam. To obtain a more optimistic estimate, we consider a pathlength of $5$ kpc where the mean value of $B_{\perp}$ is $5 \, \mu$G. For these values, the required sensitivity is $7 \, \mu$Jy/beam.

There are no publicly available observations of diffuse synchrotron emission with the required combination of sensitivity and angular resolution, however the Square Kilometre Array could be capable of achieving this combination.

\section{Conclusion and Future Work}
\label{conclusion}

We have tested the theory of synchrotron fluctuations developed by \cite{Lazarian2012} by applying it to mock observations of diffuse synchrotron emission. We have also compared statistics of the fluctuations in synchrotron intensity to the properties of the turbulence, to investigate how properties of interstellar turbulence can be deduced from observations of diffuse synchrotron emission.

We found that the NCF of the synchrotron intensity maps does depend on $\gamma$, and hence on the cosmic ray energy spectral index, in general. Thus, the common assumption that $\gamma = 2$ made in earlier papers is only valid for the subsonic and super-\alf ic, or supersonic and sub-\alf ic regimes of turbulence. However, for $\gamma$ values typically observed in the Milky Way, between $1.5$ and $2.5$, the NCF of synchrotron intensity varies little, and so independence of $\gamma$ can be assumed. We have also found that the anisotropy present in maps of diffuse synchrotron emission is well characterized by the ratio of the quadrupole to monopole moments of the normalized two-dimensional structure function of the synchrotron intensity, as predicted by \cite{Lazarian2012}. 
 
We have studied the dependence of the structure function slope and integrated quadrupole ratio modulus on $\gamma$, the sonic and \alf ic Mach numbers, and the relative orientation of the mean magnetic field relative to our line of sight for the mock synchrotron maps produced from our MHD simulations. We find that, in general, these statistics weakly depend on $\gamma$ for values of $\gamma$ typical of the Milky Way, and hence measured statistics can be compared to simulated statistics obtained for $\gamma=2$ to determine the properties of turbulence in the observed region. The exceptional case is the structure function slope for supersonic, super-\alf ic turbulence, which decreases with increasing $\gamma$. If the structure function slope is to be applied to observations of turbulence in this regime, then the value of $\gamma$ must be measured and taken into account. Although we found that the skewness and kurtosis of synchrotron intensity are not sensitive to the sonic and \alf ic Mach numbers, the structure function slope was found to be sensitive to the \alf ic Mach number, and the integrated quadrupole ratio modulus was found to be sensitive to the \alf ic Mach number, with some dependence on the sonic Mach number.

However, because the relative orientation of the mean magnetic field in the emitting region relative to the line of sight has a strong effect on the measured statistics, it is not possible to determine the sonic and \alf ic Mach numbers of the turbulence in the synchrotron emitting region unless we have prior knowledge of the magnetic field orientation, which may be obtained by observing the rotation measures of pulsars, for instance. If this prior knowledge is not available, then we may be able to analyze the polarization of the observed synchrotron emission to break these parameter degeneracies. In a future paper we will investigate how observations of the polarization gradients, and other polarization diagnostics, can help to constrain the properties of interstellar magnetized turbulence.

Finally, we estimate that a signal-to-noise ratio of $30$, an angular resolution of $21$ arc seconds (one tenth of the outer scale of turbulence) and a sensitivity of $1 \, \mu$Jy/beam are necessary to conduct a statistical analysis of the fluctuations in synchrotron intensity. These criteria may be met by the Square Kilometre Array, facilitating a study of how the magnetized turbulence in the Milky Way varies with Galactic latitude and longitude.

\begin{acknowledgments}
C.~A.~H. thanks Xiaohui Sun for his help in estimating the sensitivity required to observe synchrotron intensity fluctuations, Geraint Lewis for his supervision, and useful discussions on the real and imaginary parts of the quadrupole ratio, and Christoph Federrath for useful discussions. C.~A.~H. also thanks Sne\v zana Stanimirovi\'c, Claire Murray, Elijah Bernstein-Cooper, Bob Lindner and Brian Babler for their hospitality and useful discussions whilst at the University of Wisconsin-Madison. C.~A.~H. acknowledges financial support received via an Australian Postgraduate Award, and a Vice Chancellor's Research Scholarship awarded by the University of Sydney. The research of B.~B. is supported by the NASA Einstein Postdoctoral Fellowship. A.~L. acknowledges the NSF grant AST 1212096 and the support of the Center for Magnetic Self Organization (CMSO). A.~L. acknowledges a distinguished visitor PVE/CAPES appointment at the Physics Graduate Program of the Federal University of Rio Grande do Norte and thanks the INCT INEspao and Physics Graduate Program/UFRN for hospitality. B.~M.~G. acknowledges the support of the Australian Research Council through grant FL100100114. The Dunlap Institute for Astronomy and Astrophysics is funded through an endowment established by the David Dunlap family and the University of Toronto.
\end{acknowledgments}

\appendix
\section{Appendix A: $\gamma$ Dependence of the Structure Function and Quadrupole Ratio} \label{GamDep}

In this section we examine how changes in $\gamma$ affect the structure function and integrated quadrupole ratio modulus, for $\gamma$ values between $1$ and $4$, in increments of $0.5$. Knowledge of the $\gamma$ dependence of these statistics will aid in their interpretation. Example structure functions are shown in Figure \ref{fig11} for weak magnetic field simulations (top row), and high magnetic field simulations (bottom row). Supersonic simulations are in the left column, and subsonic simulations are in the right column. $\gamma$ values of $1, 2, 3$ and $4$ are colored blue, green, red and cyan respectively.

\begin{figure*}
\centering
\includegraphics[scale=0.7]{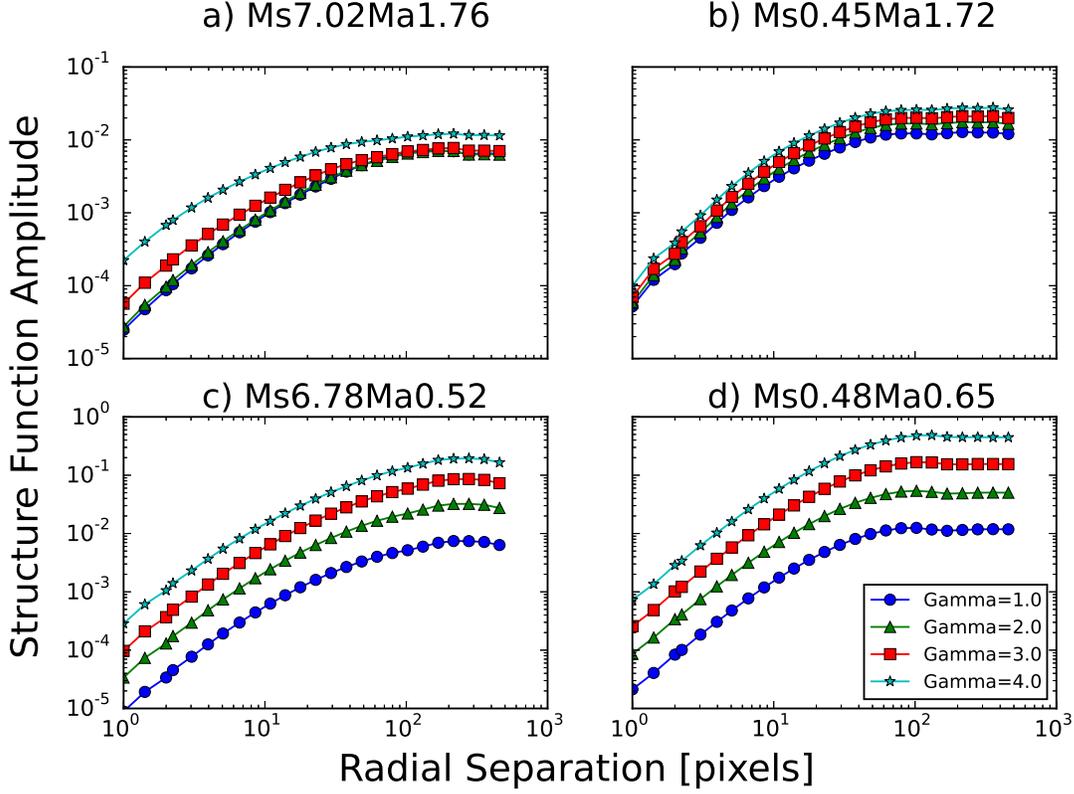}
\caption{The structure function of synchrotron intensity for the same simulations as in Figure \ref{fig2}. For each simulation, structure functions are produced for $\gamma = 1.0, 2.0, 3.0,$ and $4.0$, colored blue, green, red and cyan respectively.} \label{fig11}
\end{figure*}

We find that the structure functions for simulations with a high magnetic field have more $\gamma$ dependence in their amplitude than simulations with a low magnetic field. This is simply because the synchrotron intensity is calculated via the magnetic field strength perpendicular to our line of sight, raised to the power of $\gamma$. Thus, for high magnetic field simulations, increasing $\gamma$ leads to an increase in $I$, and hence an increase in the structure function amplitude. The structure function amplitudes of low magnetic field simulations exhibit some $\gamma$ dependence, because changing $\gamma$ changes the contrast in the observed synchrotron intensities. This changes the apparent strength of fluctuations in the synchrotron intensity, as measured by the structure function.

\begin{figure*}
\centering
\includegraphics[scale=0.7, trim=10 0 0 0, clip]{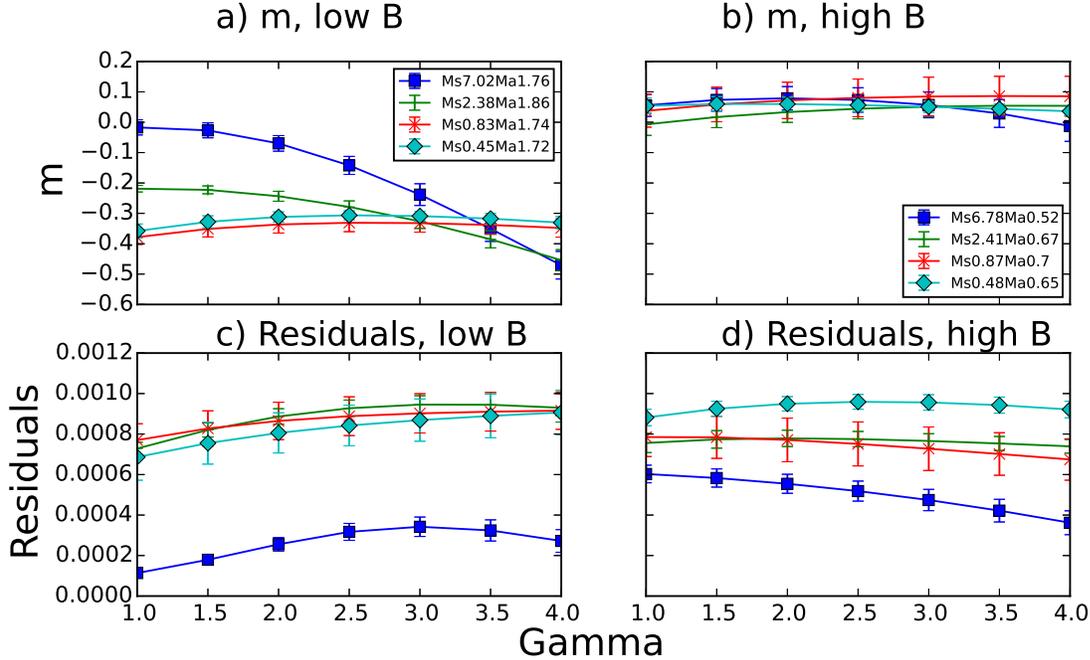}
\caption{The structure function slope minus $1$ ($m$) of synchrotron intensity as a function of $\gamma$ for a) low magnetic field and b) high magnetic field simulations. The sum of the squared residuals of the linear fit used to calculate $m$ are shown as a function of $\gamma$ for c) low magnetic field and d) high magnetic field simulations. The same simulations are used as in Figure \ref{fig8}.} \label{fig12}
\end{figure*}

We also find that the slopes of the structure functions weakly depend on $\gamma$, in general, as shown in Figure \ref{fig12}. For almost all simulations, we find that the structure function slope has little dependence on $\gamma$, particularly for $\gamma$ values typical of the Milky Way, between $1.5$ and $2.5$. Supersonic, super-\alf ic simulations are the exception, however, as for these simulations $m$ decreases with increasing $\gamma$. In Figures \ref{fig12}c) and d) we show the sum of the squared residuals of the linear fits used to calculate $m$, to demonstrate that the quality of the linear fit is similar for the simulations shown.

In Figure \ref{fig13} we show plots of the integrated quadrupole ratio modulus against $\gamma$ for simulations with an initial magnetic field strength of $0.1$ (left), and $1$ (right). For the low magnetic field simulations, we find that the integrated quadrupole ratio modulus has little dependence on $\gamma$. For the high magnetic field simulations, we find that the integrated quadrupole ratio modulus of supersonic simulations tends to decrease with increasing $\gamma$, whereas for trans- or subsonic simulations the integrated quadrupole ratio modulus tends to increase with increasing $\gamma$. This is because the anisotropic structures in the supersonic simulations are fainter than those in the trans- or subsonic simulations. Increasing the value of $\gamma$ enhances the contrast between bright and faint regions, making the anisotropic structures less prevalent for supersonic simulations, and more prevalent for subsonic simulations, which causes the integrated quadrupole ratio modulus to depend on $\gamma$.

\begin{figure*}
\centering
\includegraphics[scale=0.7, trim=25 0 0 0, clip]{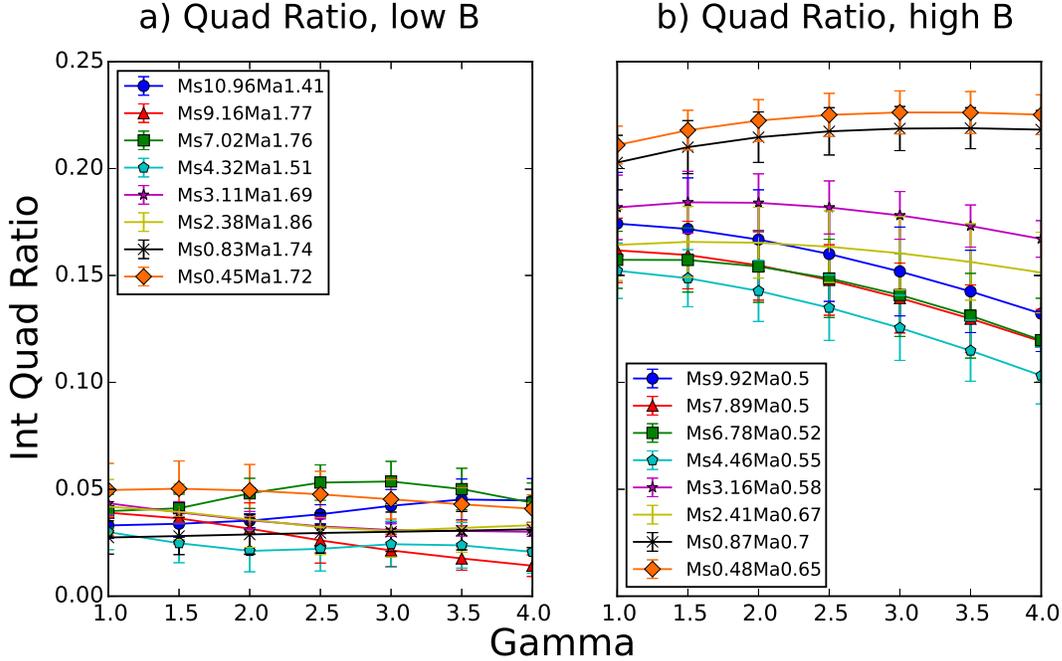}
\caption{The integrated quadrupole ratio modulus of synchrotron intensity for a) low B and b) high B simulations as a function of $\gamma$.} \label{fig13}
\end{figure*}

\section{Appendix B: Skewness and Kurtosis of Synchrotron Intensity} \label{AppSkew}
Motivated by \cite{Burkhart2009}, who found that the skewness and kurtosis of thermal electron density traced the sonic Mach number of their MHD simulations, we examined the skewness and kurtosis of our mock synchrotron intensity maps. The purpose of this study is two-fold: to investigate whether the skewness and kurtosis of synchrotron intensity are also sensitive to the sonic Mach number of our simulations, and to examine whether the skewness and kurtosis also depend on the \alf ic Mach number. 

The skewness and kurtosis statistics are the third and fourth normalized moments of a distribution respectively, and characterize how a probability distribution function deviates from a Gaussian. The skewness characterizes whether the distribution function appears skewed to the left or right, with a positive skewness implying that the distribution has an elongated tail to the right, and a negative skewness implies an elongated tail towards small values. In this study, we will calculate the biased skewness of the mock synchrotron intensity distributions, given by 

\begin{equation}
\text{Skewness} = \frac{\frac{1}{n} \sum_{i=1}^n (I_i - <I>)^3}{\left( \sqrt{\frac{1}{n} \sum_{i=1}^n (I_i - <I>)^2 }\right)^3}. \label{skewEq}
\end{equation}
In Eq. \ref{skewEq}, $n$ is the total number of pixels in the image, $<I>$ denotes the mean synchrotron intensity, and the sum is over all pixels in the image. The kurtosis characterizes whether the distribution function is flatter or more peaked than a Gaussian distribution. A negative kurtosis implies that the distribution is flatter than a Gaussian, and a positive kurtosis implies that the distribution is more peaked than a Gaussian. We calculate the biased Fisher kurtosis of the mock synchrotron intensity distributions, given by

\begin{equation}
\text{Kurtosis} = \frac{\frac{1}{n} \sum_{i=1}^n (I_i - <I>)^4}{\left( \sqrt{\frac{1}{n} \sum_{i=1}^n (I_i - <I>)^2 }\right)^4} - 3. \label{kurtEq}
\end{equation}
For each simulation in Table \ref{TabSims}, we produce plots of the skewness and kurtosis as a function of $\gamma$. Representative plots are shown in Figure \ref{fig14}, for low B simulations (left column) and high magnetic field simulations (right column).

\begin{figure*}
\centering
\includegraphics[scale=0.7, trim=30 0 0 0, clip]{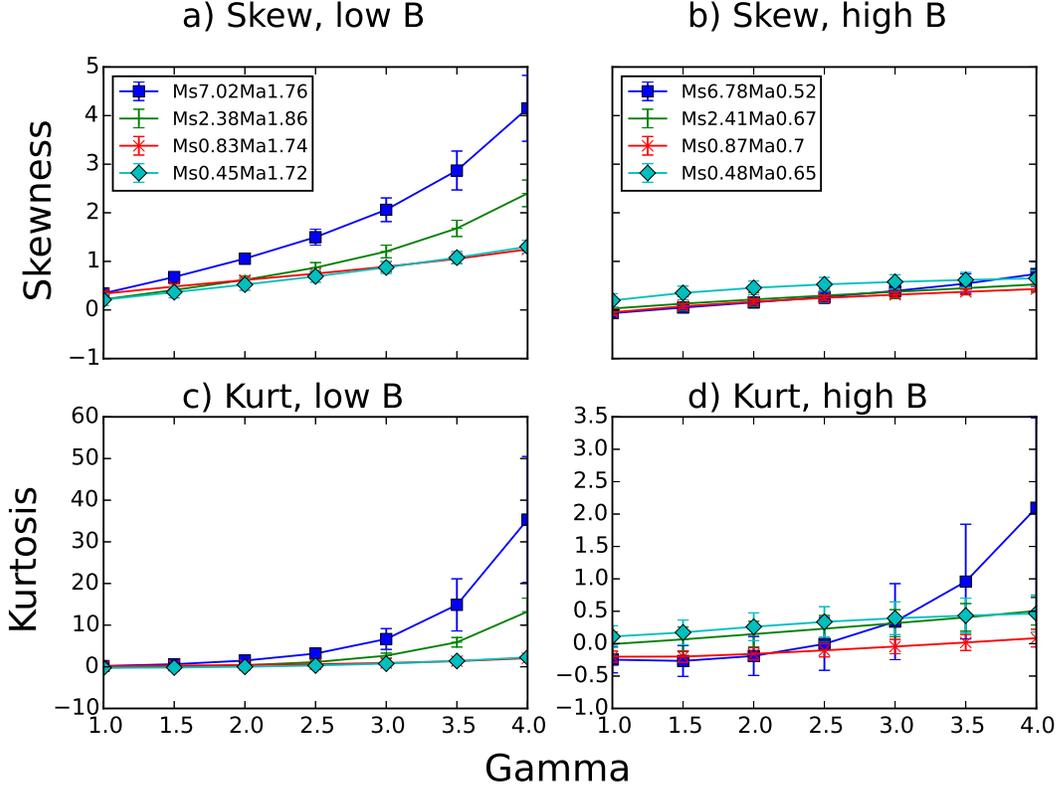}
\caption{The skewness of synchrotron intensity for a) low B and b) high B simulations as a function of $\gamma$. The kurtosis of synchrotron intensity for c) low B and d) high B simulations. The same simulations are used as in Figure \ref{fig8}.} \label{fig14}
\end{figure*}
We find that the skewness and kurtosis of most simulations depend on $\gamma$. In particular, the skewness of mock synchrotron intensity appears to increase with increasing $\gamma$, with the more supersonic, super-\alf ic simulations exhibiting greater dependence on $\gamma$. This is because the shocks that are present in the supersonic simulations compress and amplify the magnetic field. This significantly increases the synchrotron emissivity of the shocks, and produces an extended tail towards large synchrotron intensities in the synchrotron intensity distribution. Increasing the value of $\gamma$ enhances this extended tail by stretching the distribution, increasing the value of the skewness. The skewness of synchrotron intensity measured for high magnetic field simulations does not appear to have as strong a dependence on $\gamma$ as the low magnetic field simulations. This is likely because the greater magnetic pressure resists the formation of shocks, so that the tail in the synchrotron intensity distribution is not as pronounced.

As found for skewness, the kurtosis of the synchrotron intensity for low magnetic field, supersonic simulations exhibits a strong $\gamma$ dependence. This is because increasing $\gamma$ extends the tails of the synchrotron intensity distribution by stretching the distribution, without significantly broadening the synchrotron intensity distribution, so that the synchrotron intensity distribution is more peaked. Similarly, the kurtosis of synchrotron intensity for high magnetic field simulations has a stronger dependence on $\gamma$ for supersonic simulations. 

\begin{figure*}
\centering
\includegraphics[scale=0.7, trim=23 0 0 0, clip]{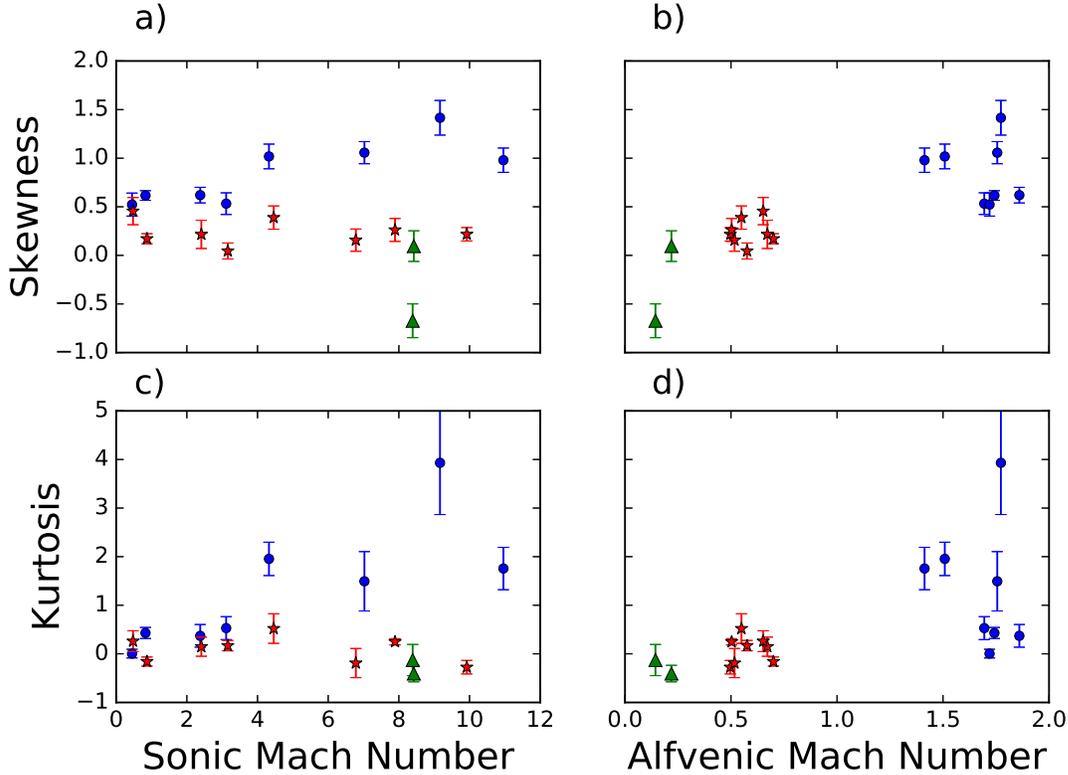}
\caption{The skewness of synchrotron intensity against a) sonic Mach number and b) \alf ic Mach number for all simulations, and $\gamma = 2$. The kurtosis of synchrotron intensity against c) sonic Mach number and d) \alf ic Mach number for all simulations, and $\gamma = 2$. Simulations with $B=0.1$ are plotted as blue circles, simulations with $B=1$ as red stars, and the remaining simulations as green triangles.} \label{fig15}
\end{figure*}

In Figure \ref{fig15}, we plot the skewness (top row) and kurtosis (bottom row) of the mock synchrotron intensity maps of all simulations produced for $\gamma = 2$ against sonic Mach number (left column) and \alf ic Mach number (right column). For both skewness and kurtosis, we find that there is a correlation with the \alf ic Mach number, where simulations with a large \alf ic Mach number tend to have a larger skewness and kurtosis. However, this correlation is not very well defined, as there is significant scatter in the skewness and kurtosis for both low and high magnetic field simulations. For example, a skewness of $0.5$ implies an \alf ic Mach number between $0.5$ and $2.0$. This covers many different regimes of turbulence, and so the skewness and kurtosis statistics are not able to distinguish these regimes. 

For the plots produced against sonic Mach number (\ref{fig15}a and \ref{fig15}c), both skewness and kurtosis display little dependence on sonic Mach number. When we partition the data into high magnetic field simulations ($B=1$, red) and low magnetic field simulations ($B=0.1$, blue), we find that for high magnetic field simulations, there is no clear dependence of the skewness or kurtosis of synchrotron intensity on the sonic Mach number. For low magnetic field simulations, we find that there is a weak correlation between the skewness and kurtosis, and the sonic Mach number. This correlation implies that large sonic Mach numbers cause an increase in the observed skewness and kurtosis. 

However, considering that simulations with large sonic Mach numbers have stronger $\gamma$ dependence than subsonic simulations, and that skewness tends to increase with $\gamma$, it is possible that skewness only appears to depend on sonic Mach number for low magnetic field simulations because of the dependence of skewness on $\gamma$. This is supported by our observation that there is no clear relationship between the skewness of synchrotron intensity maps produced using $\gamma = 1$ and sonic Mach number, for low magnetic field simulations. This finding also applies to the kurtosis of synchrotron intensity.

To summarize, we find that the skewness and kurtosis of synchrotron intensity increase with increasing \alf ic Mach number, although this is a weak trend. Due to the dependence of skewness and kurtosis on $\gamma$, we find that the skewness of synchrotron intensity does not depend on sonic Mach number for high magnetic field simulations, but does increase with sonic Mach number for low magnetic field simulations. These findings illustrate that the skewness of synchrotron intensity behaves differently to the skewness of column density. Whereas \cite{Burkhart2009} found that the skewness of column density is sensitive to the sonic Mach number, we find that the skewness of synchrotron intensity is primarily sensitive to the \alf ic Mach number, because the synchrotron intensity is sensitive to the magnetic field, but not to the thermal electron density.

\subsection{B.1. Dependence of Skewness and Kurtosis on Line of Sight} \label{AppSkewLOS}
In Figure \ref{fig16}, we plot the skewness (top row) and kurtosis (bottom row) as a function of the relative angle between the mean magnetic field and the line of sight. Low magnetic field simulations are shown on the left, and high magnetic field simulations are on the right. The same simulations are used as for Figure \ref{fig14}.

\begin{figure*}
\centering
\includegraphics[scale=0.7, trim=25 0 0 0, clip]{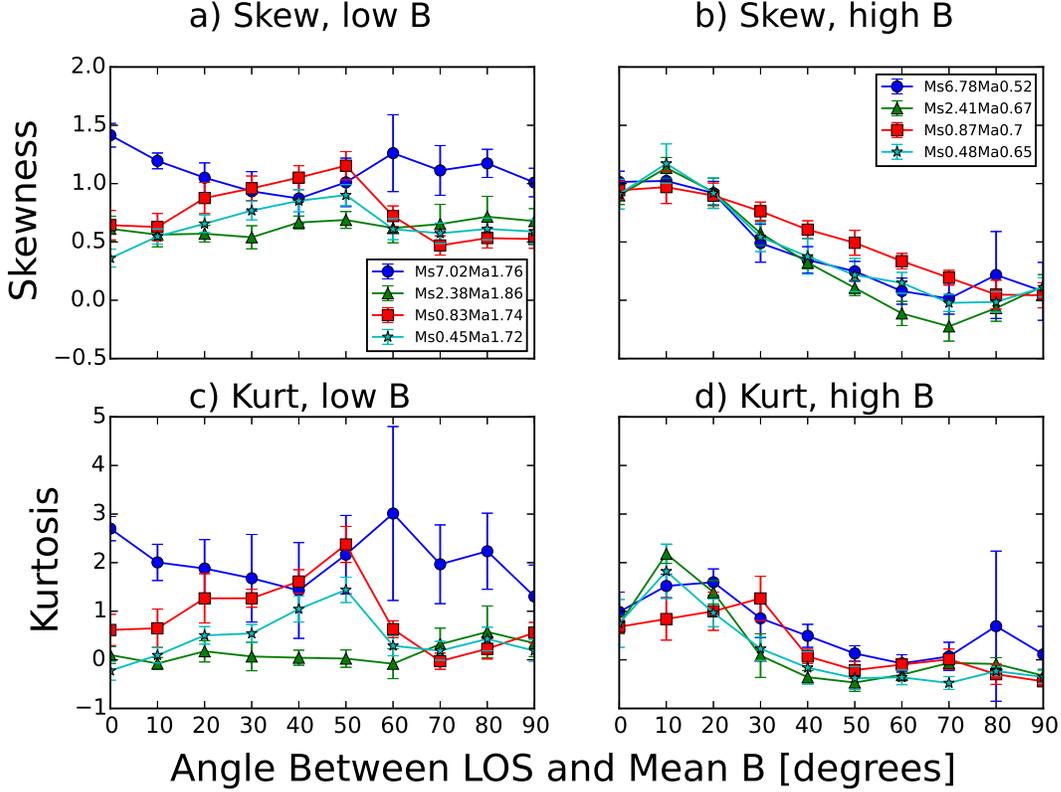}
\caption{Same as Fig. \ref{fig14}, but plotting skewness and kurtosis against the relative angle between the line of sight and the mean magnetic field, rather than $\gamma$.} \label{fig16}
\end{figure*}

We find that neither skewness nor kurtosis exhibit a clear dependence on the relative orientation of the mean magnetic field for weak magnetic field simulations. This is because there is little anisotropy in these simulations, and so the statistics should not systematically depend on the orientation of the mean magnetic field relative to the line of sight. Any variations in the skewness or kurtosis as the line of sight changes are likely caused by the particular configuration of the turbulence observed. For high magnetic field simulations, we find that both the skewness and kurtosis decrease as the line of sight becomes more perpendicular to the mean magnetic field. This occurs because the mean synchrotron intensity increases as the line of sight becomes more perpendicular to the mean magnetic field, and the fluctuations relative to the mean synchrotron intensity decrease.

Furthermore, we find that skewness only shows a weak dependence on the \alf ic Mach number of the simulations for lines of sight that are almost perpendicular to the mean magnetic field. Skewness does not exhibit any clear dependence on the sonic Mach number for any line of sight, and kurtosis does not exhibit any clear dependence on either the sonic or \alf ic Mach numbers for any line of sight. From this, we conclude that the skewness and kurtosis of synchrotron intensity are not sensitive tracers of the sonic or \alf ic Mach numbers, and that they only possess weak dependence on the relative orientation of the mean magnetic field if the field is strong.

\section{Appendix C: Dependence of Statistics on Simulation Size} \label{AppSimSize}
One way of testing whether the statistics calculated from simulations are reliable is to produce synchrotron intensity maps using a subset of the entire simulation, to examine whether small changes in the size of the mock synchrotron map produce large changes in the calculated statistics. If the statistics change drastically with small size changes, then the statistic has not converged to its true value (which would be measured with a larger simulation cube), and the statistic is not reliable.

To conduct this test, we extract a sub-cube of side length $s$ from the centre of each simulation, where $s$ varies between $312$ pixels and $512$ pixels (corresponding to the full simulation). For each sub-cube, mock synchrotron maps are produced for lines of sight along the y and z axes, and the skewness, kurtosis, structure function slope and integrated quadrupole ratio modulus of synchrotron intensity are calculated for these maps. The statistics measured for lines of sight along the y and z axes are then averaged to obtain the final data point.

In Figure \ref{fig17} we plot the skewness (top row) and kurtosis (bottom row) of the synchrotron intensity as a function of sub-cube size. Low magnetic field simulations are on the left, and high magnetic field simulations on the right. We find that the skewness and kurtosis do not change significantly with small changes in the sub-cube size for any simulation, and that there are only a couple of simulations that do not appear to have converged to their final value. We hence conclude that the skewness and kurtosis of synchrotron intensity are reliably measured in our simulations.

\begin{figure*}
\centering
\includegraphics[scale=0.7]{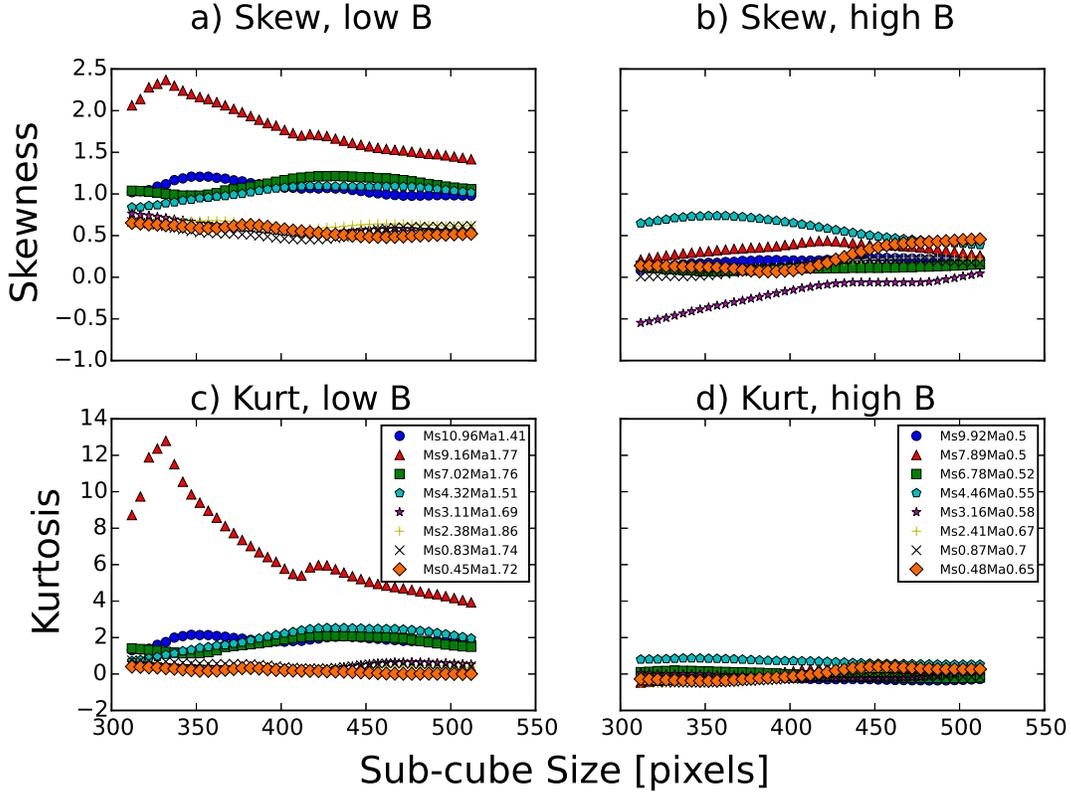}
\caption{Skewness (top row) and kurtosis (bottom row) of synchrotron intensity as a function of sub-cube size for simulations with an initial magnetic field strength of $0.1$ (left), and $1$ (right).} \label{fig17}
\end{figure*}

The structure function slope is plotted as a function of sub-cube size in Figure \ref{fig18} (top row), with low magnetic field simulations on the left, and high magnetic field simulations on the right. We find that some high magnetic field simulations do exhibit significant variations in the structure function slope for small changes in the sub-cube size, and hence the values of the structure function slope measured for these simulations are unreliable. For low magnetic field simulations, the structure function slope does not undergo significant variations, implying that the measured values of the structure function slope are reliable for these simulations. Larger simulations with strong magnetic fields must be run to obtain reliable values for the structure function slope, and results involving the structure function slope presented in this paper should hence be regarded as tentative.

\begin{figure*}
\centering
\includegraphics[scale=0.7]{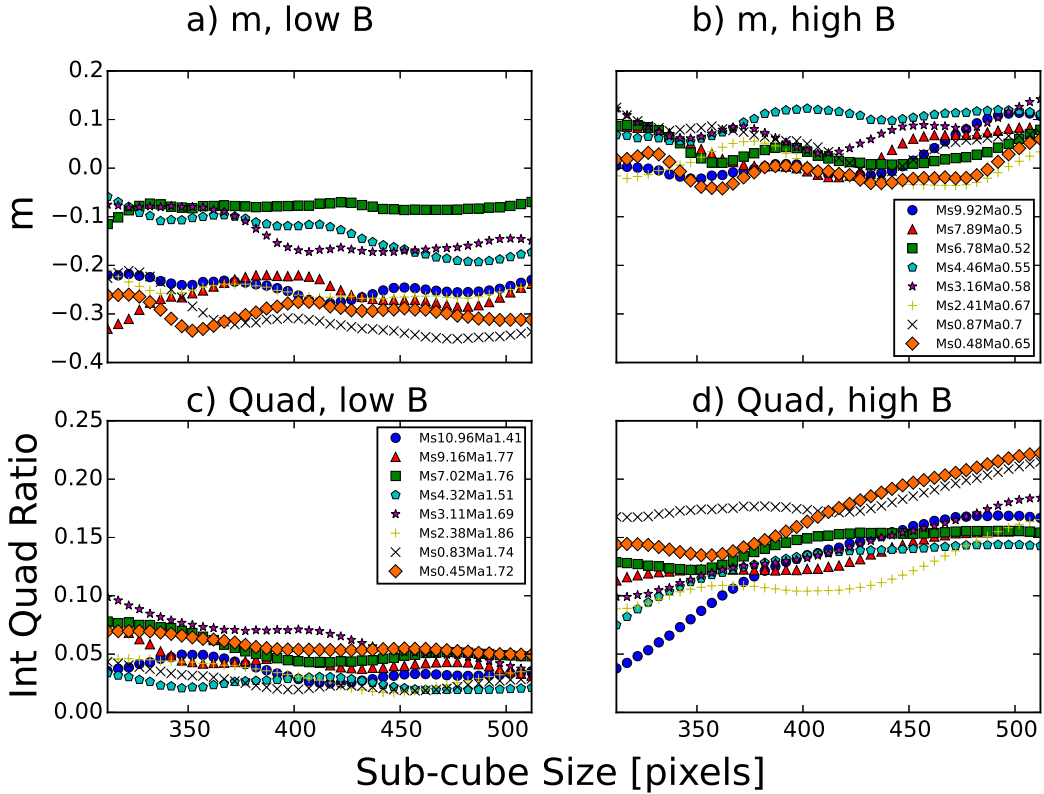}
\caption{The structure function slope minus $1$, ($m$, top row), and the integrated quadrupole ratio (bottom row) as a function of the sub-cube size. Simulations with an initial magnetic field strength of $0.1$ are on the left, and simulations with an initial magnetic field strength of $1$ are on the right.} \label{fig18}
\end{figure*}

In the bottom row of Figure \ref{fig18} we plot the integrated quadrupole ratio modulus as a function of sub-cube size. We find that simulations with a high magnetic field strength and low sonic Mach number, namely the Ms0.87Ma0.7 and Ms0.48Ma0.65 simulations, have not yet converged, or are only just beginning to converge to a final value. Otherwise, all simulations appear to plateau, and no simulations show significant changes in the integrated quadrupole ratio for small changes in size about the full simulation size. We conclude that our measurements of the integrated quadrupole ratio are reliable. 

\section{Appendix D: Implementation of Observational Effects} \label{AppObs}
\label{imp_obs}
The synchrotron maps produced by the method in Section \ref{method} are idealized, in the sense that they do not consider the spectral resolution, angular resolution, instrumental noise of the telescope, or the bandwidth used to make the final synchrotron image. In Appendix \ref{obs_effect} we will study how each of these observational effects influences the calculated statistics, with the aim of determining the angular resolution and sensitivity required to apply statistics of synchrotron fluctuations to real observations. For each effect, we consider observations being made at a central frequency of $1.4$ GHz, and use Eq. \ref{sync_inten} to scale the synchrotron intensity correspondingly. We assume $\gamma = 2$ for all observational effects studied.

To implement the effects of spectral resolution and bandwidth, we consider a spectral channel of finite width centred on $1.4$ GHz. This channel is divided up into $20$ equal sub-channels, and a synchrotron map is created for each sub-channel via Eq. \ref{mock_sync}. The synchrotron map for the full channel is then obtained by averaging over the individual sub-channel maps. For this effect, we examine $50$ different bandwidths. The values for these widths are equally spaced between $0.1$ and $300$ MHz, inclusive. The lower limit of this range is comparable to the spectral resolution of current instrumentation; for example, the Galactic ALFA Continuum Transit Survey (GALFACTS) has a spectral resolution of $0.4$ MHz \citep{Taylor2010}. The upper limit is the same as the bandwidth of the GALFACTS survey. The interpretation of these results hence changes throughout the range of spectral channel widths. Towards the lower limit, we are calculating the synchrotron intensity map that would be observed in a single spectral channel of finite spectral resolution, where the individual channel widths are far narrower than normally recorded in continuum observations. Towards the upper limit, we integrate over numerous images of synchrotron intensity to obtain a single, high signal-to-noise image.

To emulate different angular resolutions, we convolve the idealized synchrotron maps with a Gaussian kernel. We consider $20$ different standard deviations for this kernel, that start from $1$ pixel and go up to $50$ pixels in equal increments. To examine the effect of noise, we produce randomly generated $512 \times 512$ maps of Gaussian noise, with varying noise amplitudes. We determined the standard deviation of the Gaussian noise by generating linearly spaced numbers between $0.02$ and $0.5$, representing the standard deviation of the noise expressed as a fraction of the median synchrotron intensity, i.e. a noise-to-signal ratio. We then multiplied each of these numbers by the median synchrotron intensity of a map to determine the standard deviation of the noise for that map. The resultant noise map is then added onto the idealized synchrotron intensity map.

Additionally, we studied the more realistic case where both noise and angular resolution affect the observed synchrotron map by first adding a noise map to the idealized synchrotron intensity map, and then smoothing to a final resolution of $1.6$ pixels. This resolution was chosen because surveys are typically produced to have $3$ pixels across the full-width at half maximum of the telescope beam. Smoothing the synchrotron map changes how the noise contributes to the final image, and so we calculate the final noise level in the image by smoothing the idealized synchrotron intensity map to the same resolution as the final image, and subtracting the idealized map from the final map. This produces a map of the contribution of the noise to the final image, and we calculate the standard deviation of this map to determine the noise level in the smoothed image. For this study, we consider $25$ noise-to-signal ratios (relative to the original, unsmoothed image) that are linearly spaced between $0.02$ and $0.7$.

\subsection{D.1. Probing Observational Effects on Statistics} \label{obs_effect}
\subsubsection{D.1.1. Spectral Resolution and Bandwidth} \label{specres}
We find that the spectral resolution and bandwidth of the observations have no effect on the skewness, kurtosis, structure function slope or integrated quadrupole ratio modulus for bandwidths up to $300$ MHz, for all simulations. Thus, the statistics measured for a synchrotron map are very accurate, and not influenced by integrating the observed signal over the bandwidth, nor are they affected by the spectral resolution. Additionally, we have tested the impact of changing the central observing frequency between $0.8$ GHz and $2.8$ GHz, on the skewness, kurtosis, structure function slope and integrated quadrupole ratio modulus, and found that all of these statistics are independent of the observing frequency. It is possible that these statistics might change with the observing frequency, or the bandwidth, if $\gamma$ depends on frequency. If this is the case, then the statistics can be analysed in terms of how each statistic depends on $\gamma$.

\subsubsection{D.1.2. Angular Resolution}
\label{angres}
In Figure \ref{fig19} we provide an example of a mock synchrotron intensity map smoothed to a resolution of 26.8 pixels (right), using the Ms0.48Ma0.65 simulation, a line of sight perpendicular to the mean magnetic field, and $\gamma = 2$. We can convert the final resolution of this map into angular units by equating the outer scale of turbulence of the simulations (the scale at which energy is injected into the fluid motions) to the outer scale of turbulence observed in the Milky Way. \cite{Haverkorn2008} found that the outer scale of turbulence in the spiral arms of the Milky Way is a few pc, and hence we can approximate that $100$ pixels in our simulations corresponds to $3$ pc. For an emitting region that is 2 kpc away, this means that each pixel is approximately $3$ arc seconds, and thus the synchrotron map shown on the right of Figure \ref{fig19} has been smoothed to $81$ arc second resolution.

\begin{figure*}
\centering
\includegraphics[scale=0.6, trim=30 0 0 0, clip]{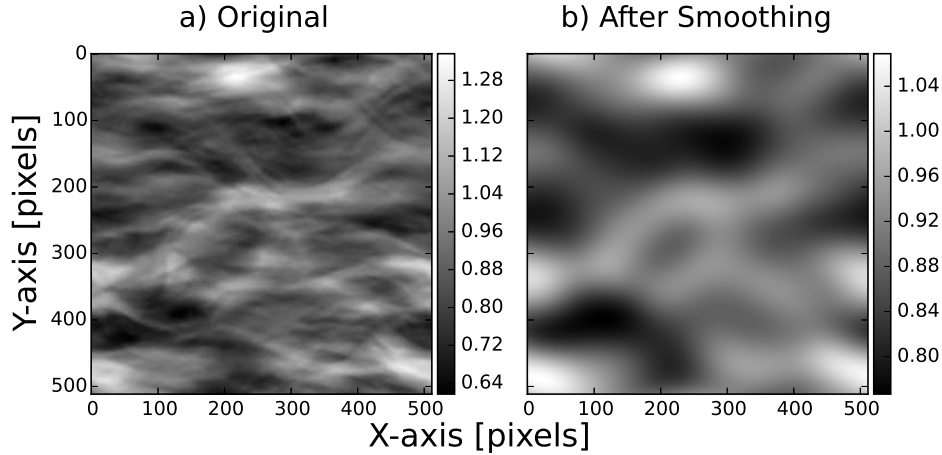}
\caption{a) The synchrotron intensity for the Ms0.48Ma0.65 simulation, for $\gamma = 2$ and a line of sight along the z-axis. b) The same synchrotron intensity image, after being smoothed with a gaussian to a resolution of 26.8 pixels.} \label{fig19}
\end{figure*}

In Figure \ref{fig20} we plot $m$ (top row) and the integrated quadrupole ratio modulus (bottom row) against the standard deviation of the Gaussian profile used to smooth the mock synchrotron intensity maps. Results for low magnetic field simulations are shown on the left, and for high magnetic field simulations on the right. We find that smoothing, and hence the angular resolution of a telescope, has a strong effect on the structure function slope, as the value of $m$ quickly deviates from its `true' value and tends towards a value of $1$ for all simulations. Thus, if the angular resolution of the telescope is too poor, above $7$ pixels ($21$ arc seconds, for an emitting region $2$ kpc away), say, then the structure function slope will no longer be able to distinguish different regimes of turbulence. We discuss the constraints this places on the angular resolution of observations hoping to make use of the structure function slope to determine properties of turbulence in Section \ref{discuss}.

\begin{figure*}
\centering
\includegraphics[scale=0.7]{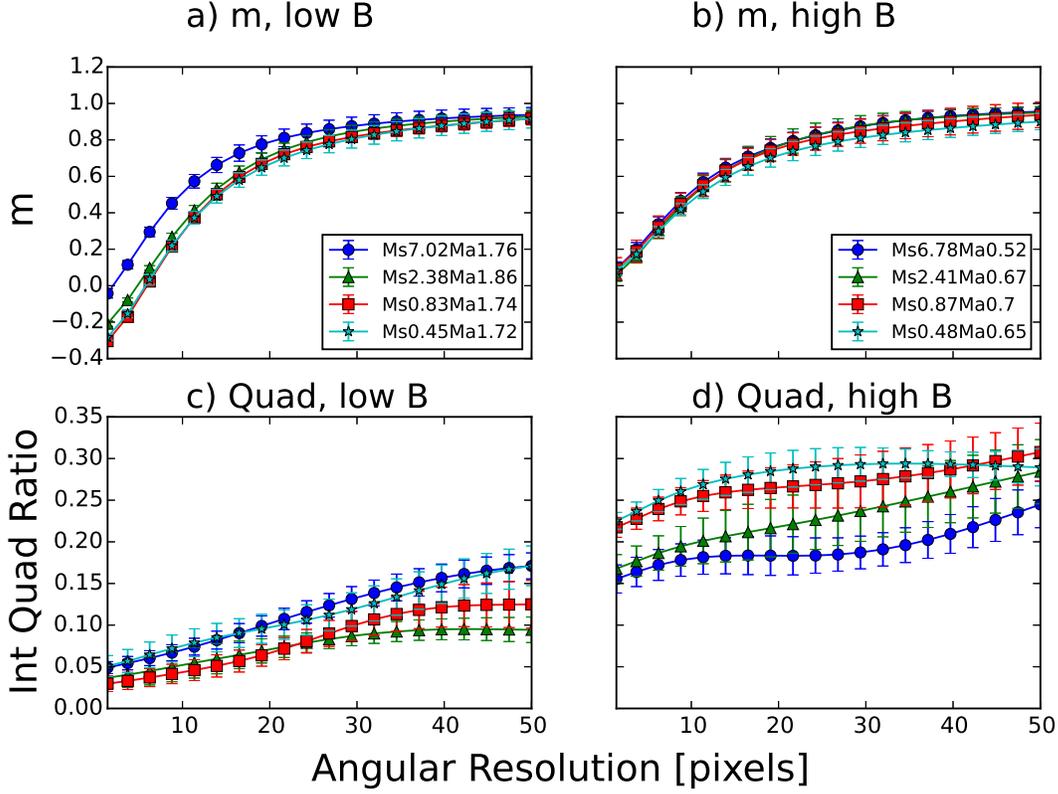}
\caption{The structure function slope minus $1$, ($m$, top row), and the integrated quadrupole ratio (bottom row) as a function of the angular resolution in pixels, for the same simulations as in Figure \ref{fig12} in the left and right columns. For an emitting region at a distance of $2$ kpc, each pixel is $3$ arc seconds across.} \label{fig20}
\end{figure*}

We find that angular resolution does not have as strong an effect on the integrated quadrupole ratio modulus as on $m$. For all simulations, there is a steady increase in the integrated quadrupole ratio modulus as the angular resolution becomes poorer. This is likely because smoothing the synchrotron intensity maps causes anisotropy that exists on scales below $20$ pixels to move into the integration range used to calculate the integrated quadrupole ratio modulus. As the quadrupole ratio tends to be larger on scales below $20$ pixels than above $20$ pixels, this leads to a gradual increase in the integrated quadrupole ratio modulus.

\subsubsection{D.1.3. Noise}
\label{noise}
In Figure \ref{fig21} we provide an example of a mock synchrotron intensity map to which noise has been added. The original mock synchrotron map produced with the Ms0.48Ma0.65 simulation, for a line of sight perpendicular to the mean magnetic field and $\gamma = 2$, is shown on the left. The mock synchrotron intensity map after Gaussian noise with a standard deviation of $27\%$ of the median synchrotron intensity has been added is shown on the right.

\begin{figure*}
\centering
\includegraphics[scale=0.6, trim=30 0 0 0, clip]{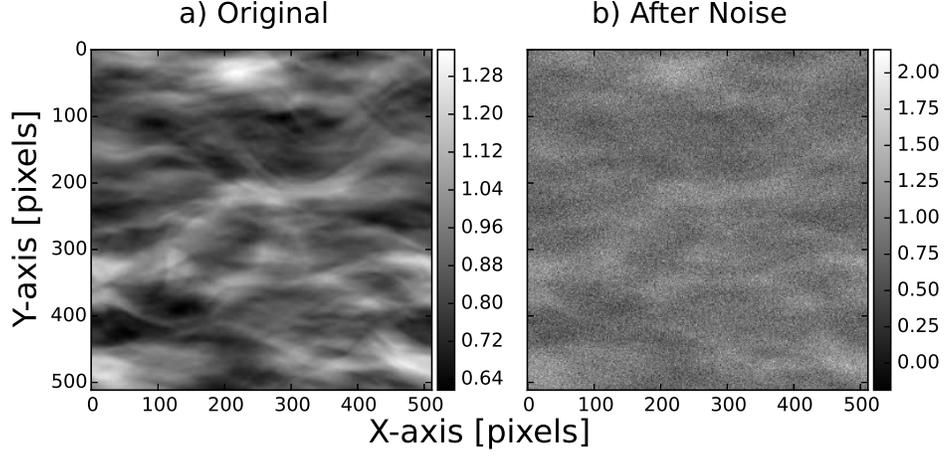}
\caption{a) The synchrotron intensity for the Ms0.48Ma0.65 simulation, for $\gamma = 2$ and a line of sight along the z-axis. b) The same synchrotron intensity image, after adding random gaussian noise with standard deviation equal to $27\%$ of the median synchrotron intensity.} \label{fig21}
\end{figure*}

We plot $m$ (top row) and the integrated quadrupole ratio modulus (bottom row) against the standard deviation of the noise (expressed as a fraction of the median synchrotron intensity of a simulation) in Figure \ref{fig22}, for the same simulations as in Figure \ref{fig20}. We find that $m$ decreases towards $-1$ as the noise level is increased, for all simulations. This is because the Gaussian noise quickly starts to dominate the statistics of the image, and so the measured structure function slope tends towards the slope measured for pure Gaussian noise, namely zero. We also find that the values of $m$ are lower for the high magnetic field simulations than the low magnetic field simulations, for a given noise level, the opposite of what was found for the original mock synchrotron intensity maps. This is likely because the median synchrotron intensity of the mock synchrotron intensity maps produced for the high magnetic field simulations is greater than that for the low magnetic field simulations, meaning that the standard deviation of the noise generated for high magnetic field simulations is larger than the standard deviation used for low magnetic field simulations.

\begin{figure*}
\centering
\includegraphics[scale=0.7]{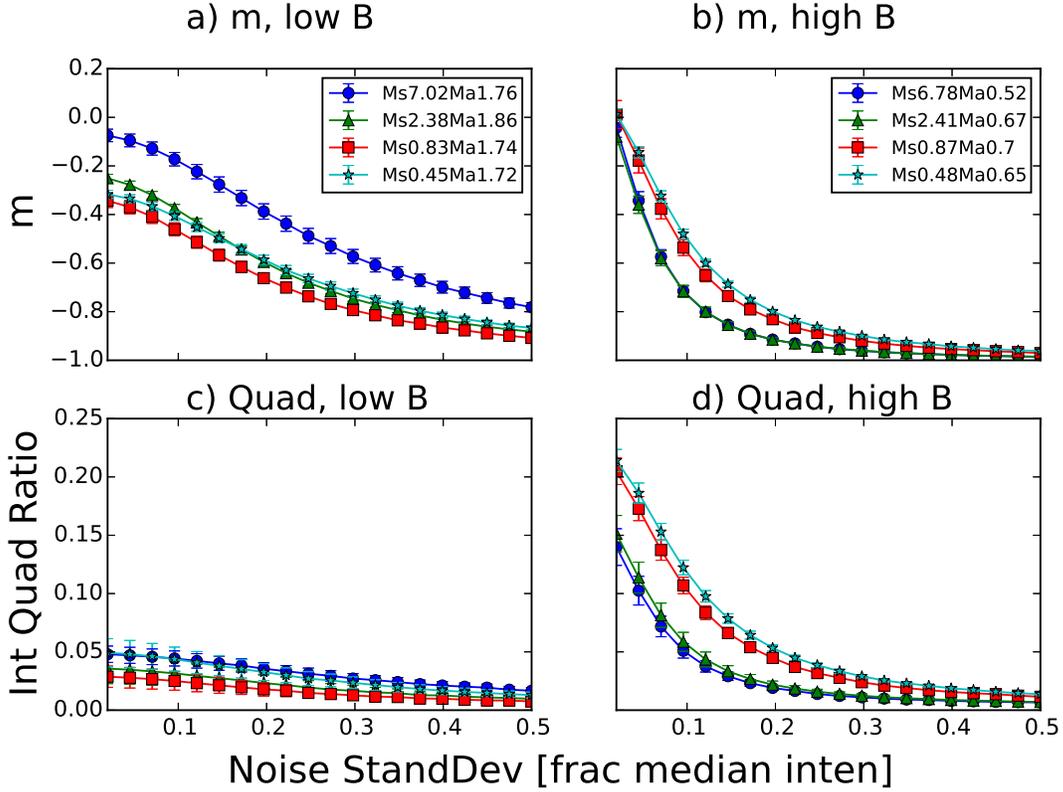}
\caption{Same as Fig. \ref{fig20}, but plotting the structure function slope minus $1$, ($m$, top row), and the integrated quadrupole ratio (bottom row) as a function of the noise level. The noise level represents the standard deviation of the noise, expressed as a fraction of the median synchrotron intensity.} \label{fig22}
\end{figure*}

We also find that the integrated quadrupole ratio modulus depends strongly on the noise level if the magnetic field strength is large, but only weakly if the magnetic field strength is small. In both cases, the integrated quadrupole ratio modulus tends towards a value of $0.01$, which we confirmed is the result obtained for pure Gaussian noise.

\subsubsection{D.1.4. Noise and Angular Resolution}
In Figure \ref{fig23} we plot $m$ (top row) and the integrated quadrupole ratio modulus (bottom row) as a function of the final noise level in the image, expressed as a fraction of the median synchrotron intensity of the final image, for a fixed resolution of $1.6$ pixels (approximately $5$ arc seconds), and the same simulations as in Figure \ref{fig20}.

\begin{figure*}
\centering
\includegraphics[scale=0.7]{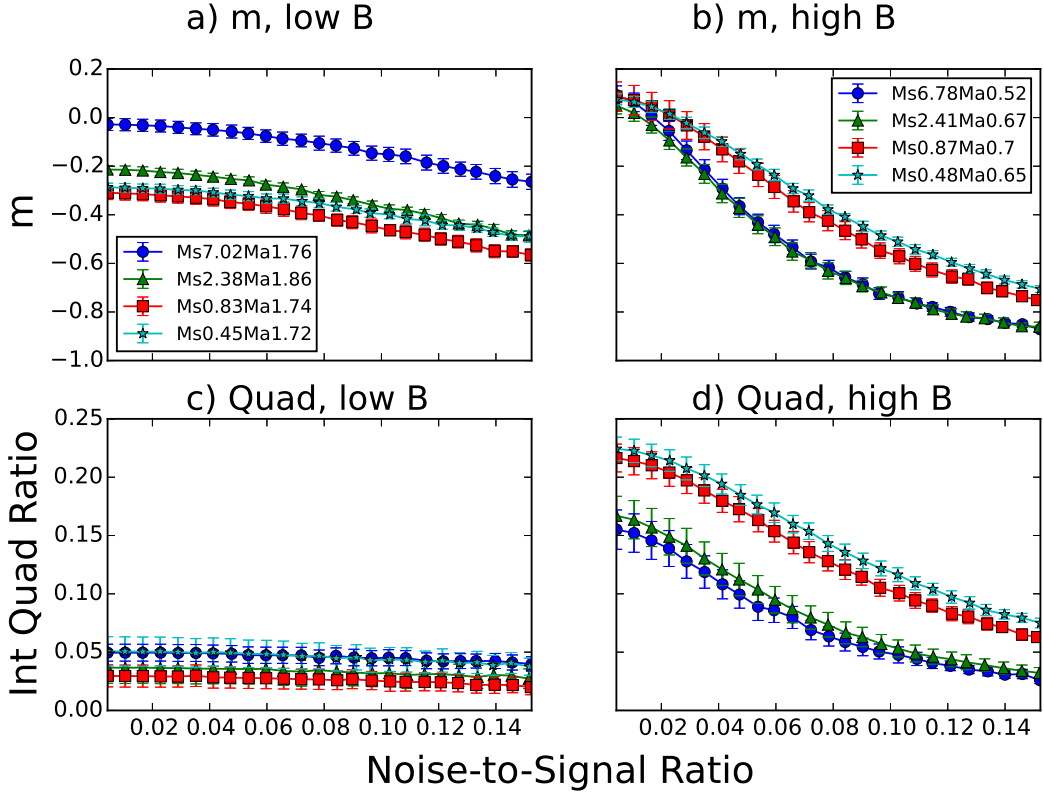}
\caption{Same as Fig. \ref{fig20}, but plotting the structure function slope minus $1$, ($m$, top row), and the integrated quadrupole ratio (bottom row) as a function of the final noise level, after the synchrotron map has been smoothed to a resolution of $1.6$ pixels. The noise level represents the standard deviation of the noise, expressed as a fraction of the median synchrotron intensity.} \label{fig23}
\end{figure*}

Firstly, we find that the noise level in the smoothed images is smaller than the initial noise level by about a factor of four, because the smoothing has averaged much of the noise. As for the case where no smoothing was performed, we find that both $m$ and the integrated quadrupole ratio modulus decrease as the noise level increases. These statistics decrease with noise level much more rapidly if the mean magnetic field perpendicular to the line of sight is large.

From these plots, we deduce that a signal-to-noise ratio of at least $20$ would be required for the measured structure function slope and integrated quadrupole ratio to be able to distinguish different regimes of turbulence. However signal-to-noise ratios above $30$ would be ideal, as this would allow for a more accurate determination of the properties of turbulence.


\bibliographystyle{../apj}
\bibliography{ref_list}

\end{document}